\documentclass[%
aps,
pre,
amsfonts,amsmath,amssymb,
notitlepage,
reprint,
10pt,
superscriptaddress,
floatfix,
nofootinbib,
letterpaper
]{revtex4-2}

\usepackage[english]{babel}
\usepackage[utf8]{inputenc}
\usepackage[T1]{fontenc}
\usepackage{mathrsfs}
\usepackage{bbm}
\usepackage{bm}
\usepackage{graphicx}
\usepackage{dcolumn}
\usepackage{dsfont}
\usepackage{enumerate}
\usepackage[dvipsnames]{xcolor}
\usepackage{float}
\usepackage{subdepth}
\usepackage{fnpct}%multiple footnotes

\usepackage[colorlinks,
linkcolor=BrickRed,
citecolor=MidnightBlue,
urlcolor=MidnightBlue,
bookmarks=true,        
bookmarksopen=true,    
bookmarksnumbered=true,
]{hyperref}

\newcounter{ls}

\newcommand*\mean[1]{\overline{#1}}
\newcommand*\tr[1]{\text{Tr}\{#1\}}

\renewcommand{\Re}{\mathrm{Re}\,}
\renewcommand{\Im}{\mathrm{Im}\,}

\begin{document}

\title{Localization and Delocalization of Quantum Trajectories in the Liouvillian Spectrum}

\author{Josef Richter}
\affiliation{Institut f\"ur Theoretische Physik, Technische Universit\"at Dresden, 01062 Dresden, Germany}

\author{Masudul Haque}
\affiliation{Institut f\"ur Theoretische Physik, Technische Universit\"at Dresden, 01062 Dresden, Germany}

\author{Lucas S\'a}
\affiliation{TCM Group, Cavendish Laboratory, University of Cambridge, JJ Thomson Avenue, Cambridge CB3 0US, UK}

\begin{abstract}
We develop an approach for understanding the dynamics of open quantum systems by analyzing individual quantum trajectories in the eigenbasis of the Liouvillian superoperator. From trajectory-eigenstate overlaps, we construct a quasiprobability distribution that characterizes the degree of localization of the trajectories in the Liouvillian eigenbasis. Contrary to the common wisdom that late-time dynamics are governed solely by the steady state and the slowest-decaying modes, we show that trajectories can remain well spread over transient eigenstates deep within the bulk of the Liouvillian spectrum even at late times. We demonstrate this explicitly using numerical simulations of interacting spin chains and bosonic systems.
Moreover, we find that the delocalization of the trajectory strongly correlates with the purity of the trajectory-averaged steady state, establishing a further link between the trajectory and ensemble pictures of open quantum dynamics.
\end{abstract}

\maketitle

\emph{Introduction.}---%
There are two complementary perspectives for describing the dynamics of open quantum systems~\cite{theory_of_open_quantum_systems}: (1) using a density matrix evolving under a suitable master equation, and (2) using ensembles of stochastic trajectories of pure-state wavefunctions. The contrast between these dual descriptions is an emerging theme in the modern understanding of non-unitary quantum dynamics.  It underlies the ongoing discussion of strong versus weak symmetries \cite{BucaProsen2012_note_on_symmetries, Albert_Jiang_PRA2014_Lindbladsymmetries, SanchezMunozBucaTindallJakschPorras_PRA2019_dissipativefreezing, LieuLundgrenAlbertGorshkov_PRL2020_symmetrybreaking, HalatiSheikhanKollath_PRR2022_BreakingStrongSymmetries, TindallJakschSanchezMunoz_SciPostPhysCore2023, WangLi_PRXQ2025_Anomaly} --- a symmetry of an open system could be preserved along every trajectory (strong) or only on average at the density matrix level (weak), and one can devise spontaneous strong-to-weak symmetry breaking transitions \cite{lee2023PRXQ,sala2024PRB, moharramipour2024PRXQ, gu2024arXiv,zhang2024arXiv, kim2024arXiv,lu2024arXiv, ando2024arXiv, chen2025PRB,sun2025PRL, orito2025PRB,lessa2025PRXQ, zhang2025PRB,liu2025CommsPhys, weinstein2025PRL,huang2025PRB, ma2025PRX, ma2025PRXQ,guo2025PRB,feng2025arXiv, sa2025SWSSB,ziereis2025arXiv}. 
The same duality is important for understanding entanglement phase transitions.  These appear in the trajectory picture but not in the ensemble-averaged picture, which is possible because nonlinear quantities such as entanglement entropies can take different values in the two. Phases distinguished by entanglement have been intensively studied in recent years in non-unitary random quantum circuits \cite{Li_2018_mipt,Skinner_2019_mipt,chan2019mipt,Li_2019_mipt,Szyniszewski2019_mipt_circuits,jian2020PRB,bao2020PRB,choi2020PRL,gullans2020_mipt,Turkeshi2020_mipt_circuits,Nahum2021_mipt_circuits,Ippoliti2021_ept_measurement_only_circuits,bao2021_circuits,Potter2022review,fisher2023review} and free fermions under weak measurement \cite{Cao_deLuca2019_entanglement_fermion_chain,Alberton2021_entanglement_transition_free_fermion_chain,Buchhold2021_mipt,fava2023PRX,Poboiko2023_free_fermions_no_mipt,Poboiko2024_mipt_free_fermion,Fava2024_mipt_free_fermion,Carisch2023_mipt,chahine2024PRB,fan2025,fan2025B,eisller2025PRA,Soares2025_entanglement_transistion_fermion_chain}.

Given the interest from multiple perspectives, the relationship between the two pictures clearly deserves detailed understanding.  
In this work, we present a route to elucidating this relationship --- we examine quantum trajectories in the eigenbasis of the Liouvillian~\cite{ferrari_2023_steadystatequantumchaosopen}.  
This leads us to challenge a common lore of open quantum systems, namely, that only the steady state and the rate at which it is approached contain relevant information, as all remaining modes decay away. Recent findings, such as non-steady-state dissipative phase transitions~\cite{prosen_2012_PT,prosen_2012_PT_2,haga2023_non_steady_state_transition,ma2025PT}, hierarchies of many-body relaxation timescales~\cite{wang2020_hierarchy,sommer2021_hierarchy,zhou2021_separation,popkov2021_spectrum,li2023_metastability,shackleton2024_spin_liquid}, transient topological modes~\cite{garcia2025_topology}, or symmetry classifications of open quantum systems~\cite{Sa_2023_SymmetryClass,kawabata2023_PRXQ,garcia2024PT} have put this view into question.  By examining the overlaps of time-evolved quantum trajectories with Liouvillian eigenstates, we illustrate and quantify the role that transient eigenstates play in late-time dynamics.

%%%%%%%%%%% BEGIN FIGURE %%%%%%%%%%%%%%%%%%%%%%%%%%%%%
\begin{figure}[h!]
    \centering
    \includegraphics[width=\linewidth]{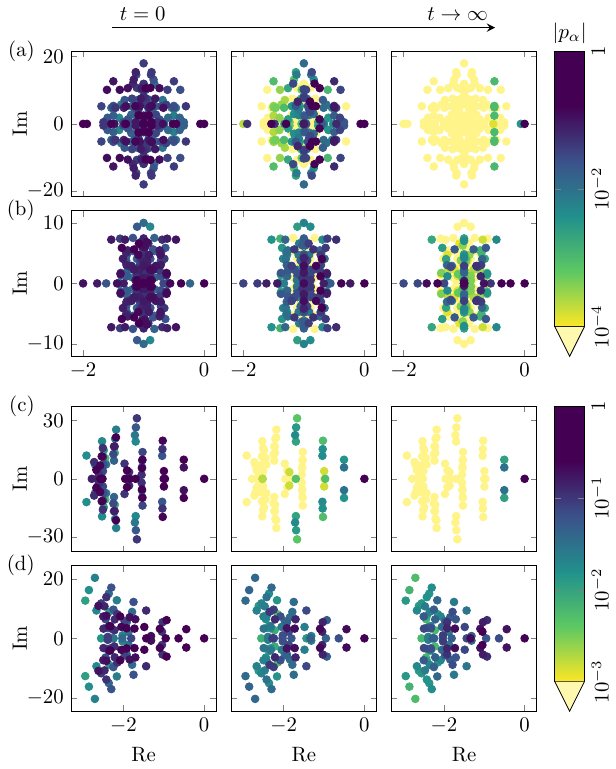}
    \caption{Absolute values of the quasiprobabilities $p_\alpha$ for each eigenvalue $\lambda_\alpha$ computed from trajectory-eigenstate overlaps for various models. Each row shows the stochastic time evolution ($t$ increases left to right) of a particular trajectory for a different model. 
    %The left column corresponds to the initial state and the right column to a trajectory at long times, while the middle one shows the state at an intermediary time.
    (a)--(b) XXZ chain, Eqs.~(\ref{eq:def:Heisenberg_XXZ_NNN}) and (\ref{eq:LB_source_sink_driving}), with $N=4$, $\Delta=0.7$, and $J'=2.0$ (a) or $J'=0.5$ (b).
    (c)--(d) Bose-Hubbard dimer, Eqs.~(\ref{eq:hamiltonian_bhdimer}) and (\ref{eq:jump_bhdimer}), with $N_c = 3$, $F=3$, and $\Delta = 2.5$ (c) or $\Delta=8$ (d).}
    \label{fig:spectra_evo}
\end{figure}
%%%%%%%%%%% END FIGURE %%%%%%%%%%%%%%%%%%%%%%%%%%%%%

Figure~\ref{fig:spectra_evo} shows the overlaps of a single quantum trajectory with the Liouvillian eigenstates, as a function of their corresponding eigenvalues $\lambda_\alpha$, for different model systems.
Starting from an initial state that is well-distributed across the full spectrum, as time progresses (left to right), the quantum trajectory can at long times overlap with either only eigenstates close to the steady state [(a) and (c)] or eigenstates in the entire spectrum [(b) and (d)], depending on the model parameters.  The present work sheds light on when quantum trajectories can be expected to be localized or delocalized in the Liouvillian spectrum, and connects this question to properties of the steady state density matrix.

Expressing the pure-state quantum trajectory in the Liouvillian eigenbasis involves some conceptual challenges stemming from the non-Hermiticity of the Liouvillian. Because the Liouvillian has distinct left and right eigenstates with non-unique normalization choices, expansion coefficients in the Liouvillian basis are not unique.  Multiple choices are also available for defining quantities like the inverse participation ratio (IPR), which is a foundational tool in studying basis (de)localization in Hermitian systems.  We provide a careful analysis of these issues, which are important beyond the present work, as overlaps appear ubiquitously in analyzing Liouvillian dynamics, e.g., in expressing the dynamics of a state in terms of the Liouvillian eigenbasis \cite{hamazaki2022_lindbladianmanybodylocalization, Zhou2023_Mpemba,  DuttaZhangHaque_PRL2025_LimitCycles, chirame2025}, in the study of the quantum Mpemba effect \cite{Lu_Raz_2017_Mpemba, Ares2025_Mpemba_review, Manikandan2021_Mpemba, Kochsiek2022_Mpemba, Ivander2023_Mpemba, carollo_2021_mpemba, chatterjee_2023_mpemba,  Nava2024_Mpemba, Wang2024_Mpemba, Liu2024_Mpemba, Graf2025_Mpemba, Furtado2025_Mpemba, Longhi2025_mpembaeffectsuper}, and non-Hermitian extensions of the eigenstate thermalization hypothesis \cite{hamazaki2022_lindbladianmanybodylocalization, cipolloni2024PRB,Hauke_PRL2025_nonHermitianETH, almeida2025_lindblad_eth,ferrari2025eth}.

\emph{The two pictures.}---The state of an open quantum system is characterized by its density matrix $\rho$, whose time evolution is generated by the Liouvillian $\mathcal{L}$, $d\rho/dt = \mathcal{L}[\rho]$.
If the time evolution is Markovian, completely positive, and trace preserving, the Liouvillian admits the Lindblad form \cite{Lindblad_1976, GKS_1976, theory_of_open_quantum_systems}
\begin{align}    
    \mathcal{L} [\rho] = -i[H, \rho] + \sum_k \gamma_k \left( L_k \rho L^\dagger_k - \frac{1}{2} \{L^\dagger_k L_k, \rho\}\right).   \label{eq:lindblad_me}
\end{align}
Here, $H$ is the Hamiltonian of the system, the jump operators $L_k$ account for the coupling to the environment, and $\gamma_k>0$ are real positive coefficients that give the coupling strength of the different decay channels $L_k$.
The dimension of $\mathcal{L}$ is $D^2$, where $D$ is the dimension of the system Hilbert space.
In the long-time limit, any initial state $\rho(0)$ will approach the steady state $\rho_{\text{ss}}$ of the Liouvillian, which in the generic case is unique. It is the eigenstate of $\mathcal{L}$ with eigenvalue zero, while all other eigenvalues have nonpositive real parts.

Quantum trajectories represent an alternative way to formulate the time evolution of the density operator as an average over pure states \cite{carmichael_open_system_approach,Dalibard1992_wavefu_approach,Molmer_93_mcwf,DumZollerRitsch1992_MC_atomic_master_eq,Gisin_1992,plenio1998_quantum_jump_approach,Daley2014}:
One can design a stochastic process that for each value of $t$ yields a random variable $|\psi_m(t)\rangle$ defined over the set of pure states, with each realization referred to as a quantum trajectory. This process is called an unravelling of the master equation \eqref{eq:lindblad_me} if the average over realizations $\{|\psi_m(t)\rangle\}_m$ converges to $\rho(t)$ with increasing sample size,
\begin{align}
\label{eq:decomposition_traj}
    \rho(t) = 
    \lim_{M\rightarrow\infty} \frac{1}{M}\sum_{m=1}^M |\psi_m(t)\rangle \langle \psi_m(t)|.
\end{align}
The unravelling of a quantum master equation into trajectories is not unique and here we consider a quantum-jump evolution of $|\psi_m(t)\rangle$ (see the End Matter for details). While all unravellings agree for quantities that depend linearly on the density matrix, they generically differ for nonlinear quantities~\cite{ wiseman1993PRA,gardiner2004_quantum_noise,wiseman2009book,theory_of_open_quantum_systems,Turkeshi2021_mipt_ising,Turkeshi2024_monitored_fermionic_chains,Pinol2024_different_unraveling,Piccitto_2024_free_fermion_different_unravellings,fan2025,fan2025B,eisller2025PRA}.

\emph{Trajectory-eigenstate overlaps.}---%
Quantum trajectories provide us with a set of pure states that define the system's state. In Hamiltonian systems, it is often useful to express quantities in the eigenbasis of the Hamiltonian; in open systems, a natural idea~\cite{ferrari_2023_steadystatequantumchaosopen} is to similarly expand the quantum trajectories in the eigenbasis of the Liouvillian.
However, because the Liouvillian is non-Hermitian, it has two distinct (left and right) sets of non-orthogonal eigenstates, 
\begin{align}
    \mathcal{L} [r_\alpha] = \lambda_\alpha r_\alpha,\quad
    \mathcal{L}^\dagger [\ell_\alpha] = \lambda_\alpha^* \ell_\alpha,
\end{align}
where $\lambda_\alpha$ is the eigenvalue corresponding to the left ($\ell_\alpha$) and right ($r_\alpha$) eigenoperators, which satisfy the biorthonormality condition
\begin{align}
    \tr{\ell^\dagger_\alpha r_\beta} = \delta_{\alpha \beta}. \label{eq:biorthonormality}
\end{align}
Assuming the Liouvillian is diagonalizable, an arbitrary density matrix $\rho$ can be expanded in either the right or the left eigenbasis:
\begin{align}
    \rho=\sum_\alpha c_{\alpha} r_\alpha=\sum_\alpha d_{\alpha} \ell_\alpha^\dagger.    \label{eq:rho_liouv_eigen_decomp}
\end{align}
Using the biorthonormality relation \eqref{eq:biorthonormality}, the complex expansion coefficients $c_\alpha$ and  $d_\alpha$ read
\begin{align}
    c_\alpha = \tr{\ell_\alpha^\dagger \rho}, \quad d_\alpha = \tr{r_\alpha \rho}.    \label{eq:expansion_coeffs}
\end{align}

This construction and our definitions of quantities of interest, such as the center of mass and the IPR, are general and valid for any quantum state $\rho$. In this letter, we will use this decomposition for the pure state  $\rho_m(t)=|\psi_m(t)\rangle\langle\psi_m(t)|$. Decomposing each trajectory $\rho_m$ according to Eq.~\eqref{eq:rho_liouv_eigen_decomp} yields time- and trajectory-dependent expansion coefficients $c_{\alpha,m}(t)$ and $d_{\alpha,m}(t)$.

\emph{Gauge freedom and quasiprobabilities.}---%
Having found, contrary to the Hamiltonian case, two distinct sets of expansion coefficients, we now clarify how they are related and if both are needed to describe physical quantities. First, we note that there is a gauge freedom in the definition of $c_\alpha$ and $d_\alpha$. Indeed, we can rescale a right eigenstate $r_\alpha$ by any complex number $k_\alpha\neq0$ without violating biorthonormality as long as the corresponding left eigenstate $\ell_\alpha$ is simultaneously rescaled by $1/k_\alpha^*$.
(Compare this with the familiar Hermitian situation, where we can rescale each eigenstate by an arbitrary phase.) 
Thus the corresponding gauge group is $\mathrm{GL}(1,\mathds{C})$, instead of $\mathrm{U}(1)$ in the Hamiltonian case.
The gauge is fixed---and, consequently, the coefficients unambiguously specified (up to a phase)---when we choose a normalization for the right (or left) eigenvectors, e.g., $\tr{r_\alpha^\dagger r_\alpha}=1$.

As any physical quantity must be independent of this gauge choice, we introduce products of the form
\begin{align}
    p_\alpha = c_\alpha d_\alpha.
\end{align}
Expressions of this form are gauge invariant, i.e., uniquely fixed by the biorthonormality relation. 
The analogous procedure in Hermitian systems is to work with probabilities instead of probability amplitudes.

We develop this analogy further and identify the quantities $p_\alpha$ as quasiprobabilities if they correspond to quantum trajectories.
The $p_\alpha$ are not probabilities in the usual sense, as they are in general complex numbers. However, they come in complex-conjugate pairs and satisfy the sum rule $\sum_\alpha p_\alpha=1$. To show the former, we note that since the Liouvillian is Hermiticity-preserving, the right (and left) eigenvectors corresponding to pairs of complex-conjugate eigenvalues are the Hermitian conjugates of each other: $\mathcal{L}[r_\alpha]=\lambda_\alpha r_\alpha\implies\mathcal{L}[r_\alpha]^\dagger=\mathcal{L}[r_\alpha^\dagger]=\lambda^*_\alpha r_\alpha^\dagger$. That is, if we denote $\lambda_{\alpha'}=\lambda_\alpha^*$, we have  $r_{\alpha'}=r_\alpha^\dagger$ and $\ell_{\alpha'}=\ell_\alpha^\dagger$, assuming a nondegenerate eigenvalue (see the End Matter for the case with degeneracies).  Hence, 
\begin{equation}
\begin{split}
c_{\alpha'}d_{\alpha'}
&=\tr{\ell_{\alpha'}^\dagger\rho}
\tr{r_{\alpha'}\rho}
\\&=\tr{\ell^\dagger_{\alpha}\rho}^*
\tr{r_{\alpha}\rho}^*
=(c_{\alpha}d_{\alpha})^*,  \label{eq:conjugate_coefficients}
\end{split}
\end{equation}
where we used the Hermiticity of $\rho$.
To establish the sum rule, by expanding it in right and left eigenstates, we write the purity of any state $\rho$ as
\begin{align}
    \label{eq:normalization_cc}
    P =\tr{\rho^2}
    &=\sum_{\alpha\beta}c_{\alpha}d_{\beta}\tr{r_\alpha\ell^\dagger_\beta}
    =\sum_{\alpha}c_\alpha d_\alpha,
\end{align}
where in the last step we used the biorthonormality relation \eqref{eq:biorthonormality}.
In particular, for pure states, such as quantum trajectories $|\psi_m(t)\rangle$, this yields the result stated above:
\begin{align}
    \sum_{\alpha}c_{\alpha,m}(t)d_{\alpha,m}(t)=1.        \label{eq:sum_rule_pure}
\end{align}

\emph{Trajectory-eigenstate overlaps in example models.}---%
To gain intuition for the quasiprobabilities $p_\alpha$, we compute them for a specific family of boundary-driven spin chains~\cite{Prosen2011_exact_ness, Prosen2012_comments_on_XXZ_chain, BucaProsen2012_note_on_symmetries, Prosen2015_spin_chain_review}; see Figs.~\ref{fig:spectra_evo}~(a) and (b). As the bulk Hamiltonian, we consider a Heisenberg XXZ chain with next-to-nearest neighbour interactions
\begin{align}
    H &= J \sum_{j=1}^{N-1} (\sigma_j^x\sigma_{j+1}^x + \sigma_j^y\sigma_{j+1}^y + \Delta \sigma_j^z\sigma_{j+1}^z) \nonumber \\
    &+ J^\prime \sum_{j=1}^{N-2} (\sigma_j^x\sigma_{j+2}^x + \sigma_j^y\sigma_{j+2}^y),  \label{eq:def:Heisenberg_XXZ_NNN}
\end{align}
with the Pauli matrices $\sigma^\alpha_j~(\alpha = x, y, z$) acting on site $j$.
We fix $J=1$ and vary $\Delta\in[-2,2]$ and $J'\in[-2,2]$.
For the Lindblad operators $L_k$ we consider source and sink driving at the boundaries of the chain,
\begin{equation}
    L_l^+ = \sigma_1^+,
    \quad
    L_r^- = \sigma_N^-,        \label{eq:LB_source_sink_driving}
\end{equation}
with the respective dissipative coupling strengths 
$\gamma_l^+=0.6$ and $\gamma_r^-=1.4$.
The chain has $N=4$ sites, hence the Hilbert space dimension is $D=2^N=16$.
We are confident that our results are generic; we demonstrate their generality by presenting analogous results for two very different bosonic models in the End Matter.  

Of particular interest is the fate of the quantum trajectories at late times, where the corresponding density matrix, given by $\lim_{t\to\infty}\rho(t)$ in Eq.~(\ref{eq:decomposition_traj}), is the steady state $\rho_{\mathrm{ss}}$.
In practice, we perform the stochastic time evolution for a set of $M$ trajectories $\{|\psi_m(t_\text{ss})\rangle\}$ up to time $t_{\mathrm{ss}}$, from which we compute the corresponding $p_{\alpha,m}(t_\text{ss})$.  (The End Matter contains further details on the numerical procedures.)
% question what is t_ss?
Figure~\ref{fig:spectra_evo} shows that the late-time distribution of $p_{\alpha}$ across the spectrum is in some models skewed toward the steady state, while in others the quasiprobabilities are distributed evenly. Consequently, eigenstates with large decay rates can still contribute appreciably to the dynamics of individual quantum trajectories even at long times.

\emph{Center of mass.}---%
To quantify the qualitative behaviour shown in Fig.~\ref{fig:spectra_evo}, we assign to the spectrum of the Liouvillian a (normalized) center of mass (CM), where the eigenvalues are weighted by the $p_\alpha$:
\begin{align}
\label{eq:CM_def}
    \text{CM}(t) := {\mean{\lambda}^{-1}} \sum_\alpha p_{\alpha}(t) \lambda_\alpha,\qquad \mean{\lambda} = \frac{1}{D^2}\sum_\alpha \lambda_\alpha.
\end{align}
Because both $\lambda_\alpha$ and $p_\alpha$ come in complex-conjugate pairs, the CM is real. Moreover, the normalization by the geometric center of the spectrum, $\mean{\lambda}$, ensures the CM is invariant under global rescalings of the Liouvillian spectrum. Finally, because $\rho_m(t)$ is a pure quantum trajectory, the CM is nonnegative (see the End Matter for a proof).

\begin{figure}[t]
    \centering
    \includegraphics[width=\linewidth]{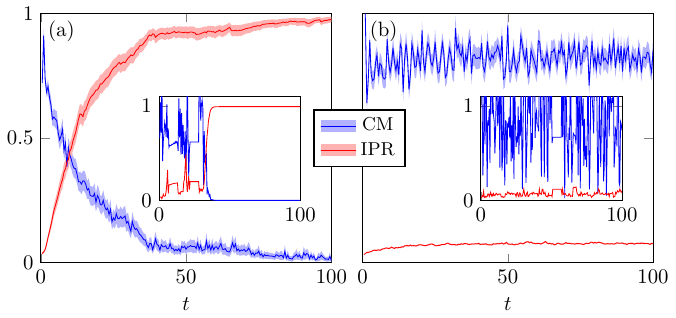}   
    \caption{Time dependence of the CM and IPR for the XXZ spin chain.  The CM and IPR are averaged over 100 trajectories in the main panels, with the shading indicating the standard deviation, while the insets show them for a single trajectory. The initial state is a state with random coefficients in the spin eigenbasis, which is fixed for all trajectories. (a) Same parameters as in Fig.~\ref{fig:spectra_evo}(a). (b) Same parameters as in Fig.~\ref{fig:spectra_evo}(b).}
    \label{fig:indicators_xxz_evo}
\end{figure}

\begin{figure}[t]
    \centering
    \includegraphics[width=\linewidth]{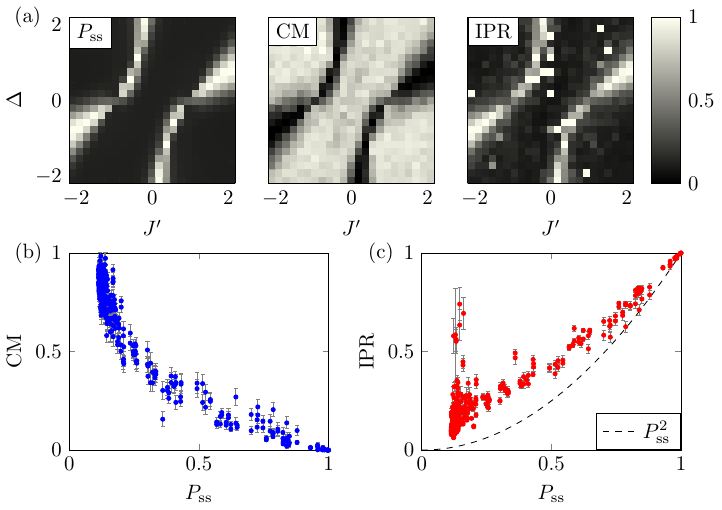}
    \caption{(a): Steady-state purity, CM, and IPR for the XXZ chain. The CM and IPR are averaged over 100 trajectories, while $P_\mathrm{ss}$ is obtained directly from the steady state. (a) Dependence of $P_\mathrm{ss}$, CM, and IPR on the parameters $J'$ and $\Delta$. The heatmaps for the three quantities show the same qualitative behaviour. (b)--(c): CM and IPR versus $P_\mathrm{ss}$. Each data point corresponds to a cell in (a). The error bars indicate the standard deviation from averaging over 100 quantum trajectories. The dashed line in (c) gives the lower bound for IPR, Eq.~(\ref{eq:P_sq_IPR}).}
    \label{fig:indicators_xxz}
\end{figure}

We present a heuristic argument to relate the late-time value of the CM to the purity $P_\text{ss}$ of the steady state $\rho_\text{ss}$.
As the quantum trajectories are pure states by definition, their overlap with the steady state tends to be smaller when $\rho_\text{ss}$ is less pure. Thus, with decreasing $P_\text{ss}$, the overlap of trajectories with other eigenstates of the Liouvillian increases, leading to a larger CM.
Conversely, in a system where $\rho_\text{ss}$ is pure ($P_\text{ss}=1$), each individual quantum trajectory has complete overlap with $\rho_\text{ss}$ in the limit $t\to \infty$ and no overlap with any other eigenstate of $\mathcal{L}$, so that $\text{CM}(t)\to0$ at large times.
We thus expect an anti-correlation between the CM and the steady state purity $P_\text{ss}$.
We verify this numerically by plotting both quantities as a function of $J'$ and $\Delta$ in the XXZ model in Fig.~\ref{fig:indicators_xxz}(a). Both quantities have the same ``phase diagram'', but with reversed heatmaps, demonstrating the argued anti-correlation. Fig.~\ref{fig:indicators_xxz}(b) shows the same anti-correlation via a scatter-plot of CM against $P_\mathrm{ss}$, each point representing one cell of Fig.~\ref{fig:indicators_xxz}(a).

\emph{Inverse participation ratio.}---%
In the Hamiltonian case, the IPR is the conventional tool to measure the degree of localization of a quantum state in a given basis. We thus seek a generalized IPR that quantifies the spread of a quantum trajectory in the Liouvillian eigenbasis.
Because the quasiprobabilities $p_\alpha$ are complex, there is an ambiguity in the definition of the IPR. The IPR defined as

\begin{align}
    \text{IPR}'(t) := \sum_\alpha \{p_\alpha(t)\}^2 = \sum_\alpha \{(\Re p_\alpha)^2-(\Im p_\alpha)^2\} \label{eq:ipr_2}
\end{align}
is unbounded; in particular, it is not necessarily positive, and hence has no obvious physical interpretation.
Instead, we define
\begin{align}
    \text{IPR}(t) := \sum_\alpha |p_\alpha(t)|^2 = \sum_\alpha \{(\Re p_\alpha)^2+(\Im p_\alpha)^2\}, \label{eq:ipr}
\end{align}
which is bounded from below by $1/D^2$ (recall that the Liouvillian has dimension $D^2)$. % \jr{$1/D^2$?}
However, as a result of the $p_\alpha$ possibly having negative real parts, this IPR is still not bounded from above and can assume arbitrarily large values.
There is no obvious ``best'' choice for the IPR that confines it to the interval $[1/D^2,1]$, but in the following we will adopt the definition~\eqref{eq:ipr}. (In the Hermitian case, both definitions~\eqref{eq:ipr_2} and \eqref{eq:ipr} reduce to the conventional IPR.)

We average the IPR over the quantum trajectories and apply the same intuition as for the CM. We expect that with a higher purity of the steady state, fewer eigenvectors contribute significantly at long times. This amounts to a larger IPR, as the $p_\alpha$ have to fulfill the sum rule~\eqref{eq:sum_rule_pure}. Almost-pure steady states restrict the significant overlaps to a small region in the spectrum, leading to an IPR close to 1. If the steady state is more mixed, the IPR decreases as more overlaps with eigenstates contribute significantly. Accordingly, the heatmap of the IPR as a function of $J'$ and $\Delta$ in Fig.~\ref{fig:indicators_xxz}(a) aligns excellently with those of the CM and $P_\mathrm{ss}$.
In Fig.~\ref{fig:indicators_xxz}(c), we plot the IPR against $P_\mathrm{ss}$ for all the cells in Fig.~\ref{fig:indicators_xxz}(a) and find that the IPR in most cases behaves as expected, showing a strong correlation with the purity, especially for large purities. The outliers reflect the fact that the IPR can take values above 1; however, the values of the IPR are typically observed to lie below 1, and thus there are few outliers.

With the strong correlation between IPR and purity in mind, we can find a lower bound for the IPR in terms of the steady state purity $P_\text{ss}$.
By inserting the representation \eqref{eq:decomposition_traj} of $\rho_\text{ss}$ into the definition of $P_\text{ss}$ and using the purity of the quantum trajectories, we obtain:
\begin{align}
    P_\text{ss} = \tr{\rho_\text{ss}^2}
    = \frac{1}{M} \sum_m \langle\psi_m|\psi_m\rangle\langle\psi_m| \rho_\text{ss} |\psi_m\rangle. \label{eq:rhoss_p0_1}
\end{align}
The steady state and the identity are the right and left eigenvectors with eigenvalue 0, respectively. Denoting the corresponding overlaps by $c_{0,m}$ and $d_{0,m}$, we thus find that
\begin{align}
    P_\text{ss} = \frac{1}{M} \sum_m c_{0,m} d_{0,m} = \mean{p_0}, \label{eq:rhoss_p0_2}
\end{align}
where the bar explicitly denotes the averaging over $|\psi_m\rangle$.
We can thus establish a lower bound on the IPR:
\begin{align}
    P_\text{ss}^2 = \mean{p_0}^2 \leq \mean{p_0^2} \leq \text{IPR}, \label{eq:P_sq_IPR}
\end{align}
with the IPR closer to saturating the bound when its dominant contribution comes from $p_0$. The bound is represented by the dashed line in Fig.~\ref{fig:indicators_xxz}. 
While the bound is exact when averaging over an infinite number of quantum trajectories, there are some apparent violations in Fig.~\ref{fig:indicators_xxz}, which we ascribe to the statistical fluctuations resulting from averaging over a finite sample of trajectories.

\emph{Conclusion.}---%
In this work, we reexamined the role of the entire Liouvillian spectrum in the late-time dynamics of open quantum systems. By analyzing individual quantum trajectories in the eigenbasis of the Liouvillian superoperator, we found that trajectories can remain highly delocalized, maintaining significant overlap with transient, rapidly decaying eigenstates deep within the bulk of the spectrum, even at times long after these modes would be expected to have decayed at the ensemble level.

To quantify the degree of localization, we constructed a quasiprobability distribution from the trajectory-eigenstate overlaps, from which we computed robust measures such as the center of mass and the inverse participation ratio of the quantum trajectory in the Liouvillian eigenbasis. This required carefully resolving the ambiguities stemming from the Liouvillian’s non-Hermiticity, including the gauge freedom of its left and right eigenvectors.

We showed that the localization of the quantum trajectories is strongly correlated with the purity of the trajectory-averaged steady state, which we verified numerically across a variety of models. Trajectory delocalization is thus not a feature of a single realization, having also direct consequences for the ensemble; this provides a further concrete link between the two complementary pictures of open quantum dynamics. It also implies that effects recently attributed to steady-state chaos~\cite{ferrari_2023_steadystatequantumchaosopen} can be understood more fundamentally in terms of this direct correlation between trajectory localization and ensemble steady-state purity.

Our paper adds to the increasing recognition of the crucial role that Liouvillian eigenvectors, and not just their eigenvalues, play in the long-time dynamics, and establishes trajectory-eigenstate overlaps as a powerful tool for our understanding of the complex interplay between the trajectory and density matrix pictures. In particular, this framework might provide a new lens for investigating phenomena where this duality is central, such as in the study of measurement-induced phase transitions. Finally, how different unravellings and the presence of strong and weak symmetries affect trajectory localization and the connection to eigenstate-eigenstate overlaps~\cite{chalker_mehlig_1998_eigenvector_statistics,mehlig_chalker_2000_eigenvector_properties} remain as interesting open questions deserving future exploration.

\bibliography{references}

%apsrev4-2.bst 2019-01-14 (MD) hand-edited version of apsrev4-1.bst
%Control: key (0)
%Control: author (8) initials jnrlst
%Control: editor formatted (1) identically to author
%Control: production of article title (0) allowed
%Control: page (0) single
%Control: year (1) truncated
%Control: production of eprint (0) enabled
\begin{thebibliography}{117}%
\makeatletter
\providecommand \@ifxundefined [1]{%
 \@ifx{#1\undefined}
}%
\providecommand \@ifnum [1]{%
 \ifnum #1\expandafter \@firstoftwo
 \else \expandafter \@secondoftwo
 \fi
}%
\providecommand \@ifx [1]{%
 \ifx #1\expandafter \@firstoftwo
 \else \expandafter \@secondoftwo
 \fi
}%
\providecommand \natexlab [1]{#1}%
\providecommand \enquote  [1]{``#1''}%
\providecommand \bibnamefont  [1]{#1}%
\providecommand \bibfnamefont [1]{#1}%
\providecommand \citenamefont [1]{#1}%
\providecommand \href@noop [0]{\@secondoftwo}%
\providecommand \href [0]{\begingroup \@sanitize@url \@href}%
\providecommand \@href[1]{\@@startlink{#1}\@@href}%
\providecommand \@@href[1]{\endgroup#1\@@endlink}%
\providecommand \@sanitize@url [0]{\catcode `\\12\catcode `\$12\catcode `\&12\catcode `\#12\catcode `\^12\catcode `\_12\catcode `\%12\relax}%
\providecommand \@@startlink[1]{}%
\providecommand \@@endlink[0]{}%
\providecommand \url  [0]{\begingroup\@sanitize@url \@url }%
\providecommand \@url [1]{\endgroup\@href {#1}{\urlprefix }}%
\providecommand \urlprefix  [0]{URL }%
\providecommand \Eprint [0]{\href }%
\providecommand \doibase [0]{https://doi.org/}%
\providecommand \selectlanguage [0]{\@gobble}%
\providecommand \bibinfo  [0]{\@secondoftwo}%
\providecommand \bibfield  [0]{\@secondoftwo}%
\providecommand \translation [1]{[#1]}%
\providecommand \BibitemOpen [0]{}%
\providecommand \bibitemStop [0]{}%
\providecommand \bibitemNoStop [0]{.\EOS\space}%
\providecommand \EOS [0]{\spacefactor3000\relax}%
\providecommand \BibitemShut  [1]{\csname bibitem#1\endcsname}%
\let\auto@bib@innerbib\@empty
%</preamble>
\bibitem [{\citenamefont {Breuer}\ and\ \citenamefont {Petruccione}(2007)}]{theory_of_open_quantum_systems}%
  \BibitemOpen
  \bibfield  {author} {\bibinfo {author} {\bibfnamefont {H.-P.}\ \bibnamefont {Breuer}}\ and\ \bibinfo {author} {\bibfnamefont {F.}~\bibnamefont {Petruccione}},\ }\href {https://doi.org/10.1093/acprof:oso/9780199213900.001.0001} {\emph {\bibinfo {title} {{The Theory of Open Quantum Systems}}}}\ (\bibinfo  {publisher} {Oxford University Press},\ \bibinfo {address} {Oxford},\ \bibinfo {year} {2007})\BibitemShut {NoStop}%
\bibitem [{\citenamefont {Buča}\ and\ \citenamefont {Prosen}(2012)}]{BucaProsen2012_note_on_symmetries}%
  \BibitemOpen
  \bibfield  {author} {\bibinfo {author} {\bibfnamefont {B.}~\bibnamefont {Buča}}\ and\ \bibinfo {author} {\bibfnamefont {T.}~\bibnamefont {Prosen}},\ }\bibfield  {title} {\bibinfo {title} {{A note on symmetry reductions of the Lindblad equation: transport in constrained open spin chains}},\ }\href {https://doi.org/10.1088/1367-2630/14/7/073007} {\bibfield  {journal} {\bibinfo  {journal} {New J. Phys.}\ }\textbf {\bibinfo {volume} {14}},\ \bibinfo {pages} {073007} (\bibinfo {year} {2012})}\BibitemShut {NoStop}%
\bibitem [{\citenamefont {Albert}\ and\ \citenamefont {Jiang}(2014)}]{Albert_Jiang_PRA2014_Lindbladsymmetries}%
  \BibitemOpen
  \bibfield  {author} {\bibinfo {author} {\bibfnamefont {V.~V.}\ \bibnamefont {Albert}}\ and\ \bibinfo {author} {\bibfnamefont {L.}~\bibnamefont {Jiang}},\ }\bibfield  {title} {\bibinfo {title} {{Symmetries and conserved quantities in {Lindblad} master equations}},\ }\href {https://doi.org/10.1103/PhysRevA.89.022118} {\bibfield  {journal} {\bibinfo  {journal} {Phys. Rev. A}\ }\textbf {\bibinfo {volume} {89}},\ \bibinfo {pages} {022118} (\bibinfo {year} {2014})}\BibitemShut {NoStop}%
\bibitem [{\citenamefont {S\'anchez Mu\~noz}\ \emph {et~al.}(2019)\citenamefont {S\'anchez Mu\~noz}, \citenamefont {Bu\v{c}a}, \citenamefont {Tindall}, \citenamefont {Gonz\'alez-Tudela}, \citenamefont {Jaksch},\ and\ \citenamefont {Porras}}]{SanchezMunozBucaTindallJakschPorras_PRA2019_dissipativefreezing}%
  \BibitemOpen
  \bibfield  {author} {\bibinfo {author} {\bibfnamefont {C.}~\bibnamefont {S\'anchez Mu\~noz}}, \bibinfo {author} {\bibfnamefont {B.}~\bibnamefont {Bu\v{c}a}}, \bibinfo {author} {\bibfnamefont {J.}~\bibnamefont {Tindall}}, \bibinfo {author} {\bibfnamefont {A.}~\bibnamefont {Gonz\'alez-Tudela}}, \bibinfo {author} {\bibfnamefont {D.}~\bibnamefont {Jaksch}},\ and\ \bibinfo {author} {\bibfnamefont {D.}~\bibnamefont {Porras}},\ }\bibfield  {title} {\bibinfo {title} {{Symmetries and conservation laws in quantum trajectories: Dissipative freezing}},\ }\href {https://doi.org/10.1103/PhysRevA.100.042113} {\bibfield  {journal} {\bibinfo  {journal} {Phys. Rev. A}\ }\textbf {\bibinfo {volume} {100}},\ \bibinfo {pages} {042113} (\bibinfo {year} {2019})}\BibitemShut {NoStop}%
\bibitem [{\citenamefont {Lieu}\ \emph {et~al.}(2020)\citenamefont {Lieu}, \citenamefont {Belyansky}, \citenamefont {Young}, \citenamefont {Lundgren}, \citenamefont {Albert},\ and\ \citenamefont {Gorshkov}}]{LieuLundgrenAlbertGorshkov_PRL2020_symmetrybreaking}%
  \BibitemOpen
  \bibfield  {author} {\bibinfo {author} {\bibfnamefont {S.}~\bibnamefont {Lieu}}, \bibinfo {author} {\bibfnamefont {R.}~\bibnamefont {Belyansky}}, \bibinfo {author} {\bibfnamefont {J.~T.}\ \bibnamefont {Young}}, \bibinfo {author} {\bibfnamefont {R.}~\bibnamefont {Lundgren}}, \bibinfo {author} {\bibfnamefont {V.~V.}\ \bibnamefont {Albert}},\ and\ \bibinfo {author} {\bibfnamefont {A.~V.}\ \bibnamefont {Gorshkov}},\ }\bibfield  {title} {\bibinfo {title} {{Symmetry Breaking and Error Correction in Open Quantum Systems}},\ }\href {https://doi.org/10.1103/PhysRevLett.125.240405} {\bibfield  {journal} {\bibinfo  {journal} {Phys. Rev. Lett.}\ }\textbf {\bibinfo {volume} {125}},\ \bibinfo {pages} {240405} (\bibinfo {year} {2020})}\BibitemShut {NoStop}%
\bibitem [{\citenamefont {Halati}\ \emph {et~al.}(2022)\citenamefont {Halati}, \citenamefont {Sheikhan},\ and\ \citenamefont {Kollath}}]{HalatiSheikhanKollath_PRR2022_BreakingStrongSymmetries}%
  \BibitemOpen
  \bibfield  {author} {\bibinfo {author} {\bibfnamefont {C.-M.}\ \bibnamefont {Halati}}, \bibinfo {author} {\bibfnamefont {A.}~\bibnamefont {Sheikhan}},\ and\ \bibinfo {author} {\bibfnamefont {C.}~\bibnamefont {Kollath}},\ }\bibfield  {title} {\bibinfo {title} {{Breaking strong symmetries in dissipative quantum systems: Bosonic atoms coupled to a cavity}},\ }\href {https://doi.org/10.1103/PhysRevResearch.4.L012015} {\bibfield  {journal} {\bibinfo  {journal} {Phys. Rev. Res.}\ }\textbf {\bibinfo {volume} {4}},\ \bibinfo {pages} {L012015} (\bibinfo {year} {2022})}\BibitemShut {NoStop}%
\bibitem [{\citenamefont {Tindall}\ \emph {et~al.}(2023)\citenamefont {Tindall}, \citenamefont {Jaksch},\ and\ \citenamefont {Muñoz}}]{TindallJakschSanchezMunoz_SciPostPhysCore2023}%
  \BibitemOpen
  \bibfield  {author} {\bibinfo {author} {\bibfnamefont {J.}~\bibnamefont {Tindall}}, \bibinfo {author} {\bibfnamefont {D.}~\bibnamefont {Jaksch}},\ and\ \bibinfo {author} {\bibfnamefont {C.~S.}\ \bibnamefont {Muñoz}},\ }\bibfield  {title} {\bibinfo {title} {{On the generality of symmetry breaking and dissipative freezing in quantum trajectories}},\ }\href {https://doi.org/10.21468/SciPostPhysCore.6.1.004} {\bibfield  {journal} {\bibinfo  {journal} {SciPost Phys. Core}\ }\textbf {\bibinfo {volume} {6}},\ \bibinfo {pages} {004} (\bibinfo {year} {2023})}\BibitemShut {NoStop}%
\bibitem [{\citenamefont {Wang}\ and\ \citenamefont {Li}(2025)}]{WangLi_PRXQ2025_Anomaly}%
  \BibitemOpen
  \bibfield  {author} {\bibinfo {author} {\bibfnamefont {Z.}~\bibnamefont {Wang}}\ and\ \bibinfo {author} {\bibfnamefont {L.}~\bibnamefont {Li}},\ }\bibfield  {title} {\bibinfo {title} {{Anomaly in Open Quantum Systems and its Implications on Mixed-State Quantum Phases}},\ }\href {https://doi.org/10.1103/PRXQuantum.6.010347} {\bibfield  {journal} {\bibinfo  {journal} {PRX Quantum}\ }\textbf {\bibinfo {volume} {6}},\ \bibinfo {pages} {010347} (\bibinfo {year} {2025})}\BibitemShut {NoStop}%
\bibitem [{\citenamefont {Lee}\ \emph {et~al.}(2023)\citenamefont {Lee}, \citenamefont {Jian},\ and\ \citenamefont {Xu}}]{lee2023PRXQ}%
  \BibitemOpen
  \bibfield  {author} {\bibinfo {author} {\bibfnamefont {J.~Y.}\ \bibnamefont {Lee}}, \bibinfo {author} {\bibfnamefont {C.-M.}\ \bibnamefont {Jian}},\ and\ \bibinfo {author} {\bibfnamefont {C.}~\bibnamefont {Xu}},\ }\bibfield  {title} {\bibinfo {title} {{Quantum Criticality Under Decoherence or Weak Measurement}},\ }\href {https://doi.org/10.1103/PRXQuantum.4.030317} {\bibfield  {journal} {\bibinfo  {journal} {PRX Quantum}\ }\textbf {\bibinfo {volume} {4}},\ \bibinfo {pages} {030317} (\bibinfo {year} {2023})}\BibitemShut {NoStop}%
\bibitem [{\citenamefont {Sala}\ \emph {et~al.}(2024)\citenamefont {Sala}, \citenamefont {Gopalakrishnan}, \citenamefont {Oshikawa},\ and\ \citenamefont {You}}]{sala2024PRB}%
  \BibitemOpen
  \bibfield  {author} {\bibinfo {author} {\bibfnamefont {P.}~\bibnamefont {Sala}}, \bibinfo {author} {\bibfnamefont {S.}~\bibnamefont {Gopalakrishnan}}, \bibinfo {author} {\bibfnamefont {M.}~\bibnamefont {Oshikawa}},\ and\ \bibinfo {author} {\bibfnamefont {Y.}~\bibnamefont {You}},\ }\bibfield  {title} {\bibinfo {title} {{Spontaneous strong symmetry breaking in open systems: Purification perspective}},\ }\href {https://doi.org/10.1103/PhysRevB.110.155150} {\bibfield  {journal} {\bibinfo  {journal} {Phys. Rev. B}\ }\textbf {\bibinfo {volume} {110}},\ \bibinfo {pages} {155150} (\bibinfo {year} {2024})}\BibitemShut {NoStop}%
\bibitem [{\citenamefont {Moharramipour}\ \emph {et~al.}(2024)\citenamefont {Moharramipour}, \citenamefont {Lessa}, \citenamefont {Wang}, \citenamefont {Hsieh},\ and\ \citenamefont {Sahu}}]{moharramipour2024PRXQ}%
  \BibitemOpen
  \bibfield  {author} {\bibinfo {author} {\bibfnamefont {A.}~\bibnamefont {Moharramipour}}, \bibinfo {author} {\bibfnamefont {L.~A.}\ \bibnamefont {Lessa}}, \bibinfo {author} {\bibfnamefont {C.}~\bibnamefont {Wang}}, \bibinfo {author} {\bibfnamefont {T.~H.}\ \bibnamefont {Hsieh}},\ and\ \bibinfo {author} {\bibfnamefont {S.}~\bibnamefont {Sahu}},\ }\bibfield  {title} {\bibinfo {title} {{Symmetry-Enforced Entanglement in Maximally Mixed States}},\ }\href {https://doi.org/10.1103/PRXQuantum.5.040336} {\bibfield  {journal} {\bibinfo  {journal} {PRX Quantum}\ }\textbf {\bibinfo {volume} {5}},\ \bibinfo {pages} {040336} (\bibinfo {year} {2024})}\BibitemShut {NoStop}%
\bibitem [{\citenamefont {Gu}\ \emph {et~al.}(2024)\citenamefont {Gu}, \citenamefont {Wang},\ and\ \citenamefont {Wang}}]{gu2024arXiv}%
  \BibitemOpen
  \bibfield  {author} {\bibinfo {author} {\bibfnamefont {D.}~\bibnamefont {Gu}}, \bibinfo {author} {\bibfnamefont {Z.}~\bibnamefont {Wang}},\ and\ \bibinfo {author} {\bibfnamefont {Z.}~\bibnamefont {Wang}},\ }\bibfield  {title} {\bibinfo {title} {{Spontaneous symmetry breaking in open quantum systems: strong, weak, and strong-to-weak}},\ }\href {https://arxiv.org/abs/2406.19381} {\bibfield  {journal} {\bibinfo  {journal} {arXiv:2406.19381}\ } (\bibinfo {year} {2024})}\BibitemShut {NoStop}%
\bibitem [{\citenamefont {Zhang}\ \emph {et~al.}(2024)\citenamefont {Zhang}, \citenamefont {Xu},\ and\ \citenamefont {Xu}}]{zhang2024arXiv}%
  \BibitemOpen
  \bibfield  {author} {\bibinfo {author} {\bibfnamefont {J.-H.}\ \bibnamefont {Zhang}}, \bibinfo {author} {\bibfnamefont {C.}~\bibnamefont {Xu}},\ and\ \bibinfo {author} {\bibfnamefont {Y.}~\bibnamefont {Xu}},\ }\bibfield  {title} {\bibinfo {title} {{Fluctuation-dissipation theorem and information geometry in open quantum systems}},\ }\href {https://arxiv.org/abs/2409.18944} {\bibfield  {journal} {\bibinfo  {journal} {arXiv:2409.18944}\ } (\bibinfo {year} {2024})}\BibitemShut {NoStop}%
\bibitem [{\citenamefont {Kim}\ \emph {et~al.}(2024)\citenamefont {Kim}, \citenamefont {Altman},\ and\ \citenamefont {Lee}}]{kim2024arXiv}%
  \BibitemOpen
  \bibfield  {author} {\bibinfo {author} {\bibfnamefont {J.}~\bibnamefont {Kim}}, \bibinfo {author} {\bibfnamefont {E.}~\bibnamefont {Altman}},\ and\ \bibinfo {author} {\bibfnamefont {J.~Y.}\ \bibnamefont {Lee}},\ }\bibfield  {title} {\bibinfo {title} {{Error Threshold of SYK Codes from Strong-to-Weak Parity Symmetry Breaking}},\ }\href {https://arxiv.org/abs/2410.24225} {\bibfield  {journal} {\bibinfo  {journal} {arXiv:2410.24225}\ } (\bibinfo {year} {2024})}\BibitemShut {NoStop}%
\bibitem [{\citenamefont {Lu}\ \emph {et~al.}(2024)\citenamefont {Lu}, \citenamefont {Zhu},\ and\ \citenamefont {Lu}}]{lu2024arXiv}%
  \BibitemOpen
  \bibfield  {author} {\bibinfo {author} {\bibfnamefont {S.}~\bibnamefont {Lu}}, \bibinfo {author} {\bibfnamefont {P.}~\bibnamefont {Zhu}},\ and\ \bibinfo {author} {\bibfnamefont {Y.-M.}\ \bibnamefont {Lu}},\ }\bibfield  {title} {\bibinfo {title} {{Bilayer construction for mixed state phenomena with strong, weak symmetries and symmetry breakings}},\ }\href {https://arxiv.org/abs/2411.07174} {\bibfield  {journal} {\bibinfo  {journal} {arXiv:2411.07174}\ } (\bibinfo {year} {2024})}\BibitemShut {NoStop}%
\bibitem [{\citenamefont {Ando}\ \emph {et~al.}(2024)\citenamefont {Ando}, \citenamefont {Ryu},\ and\ \citenamefont {Watanabe}}]{ando2024arXiv}%
  \BibitemOpen
  \bibfield  {author} {\bibinfo {author} {\bibfnamefont {T.}~\bibnamefont {Ando}}, \bibinfo {author} {\bibfnamefont {S.}~\bibnamefont {Ryu}},\ and\ \bibinfo {author} {\bibfnamefont {M.}~\bibnamefont {Watanabe}},\ }\bibfield  {title} {\bibinfo {title} {{Gauge theory and mixed state criticality}},\ }\href {https://arxiv.org/abs/2411.04360} {\bibfield  {journal} {\bibinfo  {journal} {arXiv:2411.04360}\ } (\bibinfo {year} {2024})}\BibitemShut {NoStop}%
\bibitem [{\citenamefont {Chen}\ \emph {et~al.}(2025)\citenamefont {Chen}, \citenamefont {Sun},\ and\ \citenamefont {Zhang}}]{chen2025PRB}%
  \BibitemOpen
  \bibfield  {author} {\bibinfo {author} {\bibfnamefont {L.}~\bibnamefont {Chen}}, \bibinfo {author} {\bibfnamefont {N.}~\bibnamefont {Sun}},\ and\ \bibinfo {author} {\bibfnamefont {P.}~\bibnamefont {Zhang}},\ }\bibfield  {title} {\bibinfo {title} {{Strong-to-weak symmetry breaking and entanglement transitions}},\ }\href {https://doi.org/10.1103/PhysRevB.111.L060304} {\bibfield  {journal} {\bibinfo  {journal} {Phys. Rev. B}\ }\textbf {\bibinfo {volume} {111}},\ \bibinfo {pages} {L060304} (\bibinfo {year} {2025})}\BibitemShut {NoStop}%
\bibitem [{\citenamefont {Sun}\ \emph {et~al.}(2025)\citenamefont {Sun}, \citenamefont {Zhang},\ and\ \citenamefont {Feng}}]{sun2025PRL}%
  \BibitemOpen
  \bibfield  {author} {\bibinfo {author} {\bibfnamefont {N.}~\bibnamefont {Sun}}, \bibinfo {author} {\bibfnamefont {P.}~\bibnamefont {Zhang}},\ and\ \bibinfo {author} {\bibfnamefont {L.}~\bibnamefont {Feng}},\ }\bibfield  {title} {\bibinfo {title} {{Scheme to Detect the Strong-to-Weak Symmetry Breaking via Randomized Measurements}},\ }\href {https://doi.org/10.1103/7p5x-7yqb} {\bibfield  {journal} {\bibinfo  {journal} {Phys. Rev. Lett.}\ }\textbf {\bibinfo {volume} {135}},\ \bibinfo {pages} {090403} (\bibinfo {year} {2025})}\BibitemShut {NoStop}%
\bibitem [{\citenamefont {Orito}\ \emph {et~al.}(2025)\citenamefont {Orito}, \citenamefont {Kuno},\ and\ \citenamefont {Ichinose}}]{orito2025PRB}%
  \BibitemOpen
  \bibfield  {author} {\bibinfo {author} {\bibfnamefont {T.}~\bibnamefont {Orito}}, \bibinfo {author} {\bibfnamefont {Y.}~\bibnamefont {Kuno}},\ and\ \bibinfo {author} {\bibfnamefont {I.}~\bibnamefont {Ichinose}},\ }\bibfield  {title} {\bibinfo {title} {{Strong and weak symmetries and their spontaneous symmetry breaking in mixed states emerging from the quantum Ising model under multiple decoherence}},\ }\href {https://doi.org/10.1103/PhysRevB.111.054106} {\bibfield  {journal} {\bibinfo  {journal} {Phys. Rev. B}\ }\textbf {\bibinfo {volume} {111}},\ \bibinfo {pages} {054106} (\bibinfo {year} {2025})}\BibitemShut {NoStop}%
\bibitem [{\citenamefont {Lessa}\ \emph {et~al.}(2025)\citenamefont {Lessa}, \citenamefont {Ma}, \citenamefont {Zhang}, \citenamefont {Bi}, \citenamefont {Cheng},\ and\ \citenamefont {Wang}}]{lessa2025PRXQ}%
  \BibitemOpen
  \bibfield  {author} {\bibinfo {author} {\bibfnamefont {L.~A.}\ \bibnamefont {Lessa}}, \bibinfo {author} {\bibfnamefont {R.}~\bibnamefont {Ma}}, \bibinfo {author} {\bibfnamefont {J.-H.}\ \bibnamefont {Zhang}}, \bibinfo {author} {\bibfnamefont {Z.}~\bibnamefont {Bi}}, \bibinfo {author} {\bibfnamefont {M.}~\bibnamefont {Cheng}},\ and\ \bibinfo {author} {\bibfnamefont {C.}~\bibnamefont {Wang}},\ }\bibfield  {title} {\bibinfo {title} {{Strong-to-Weak Spontaneous Symmetry Breaking in Mixed Quantum States}},\ }\href {https://doi.org/10.1103/PRXQuantum.6.010344} {\bibfield  {journal} {\bibinfo  {journal} {PRX Quantum}\ }\textbf {\bibinfo {volume} {6}},\ \bibinfo {pages} {010344} (\bibinfo {year} {2025})}\BibitemShut {NoStop}%
\bibitem [{\citenamefont {Zhang}\ \emph {et~al.}(2025)\citenamefont {Zhang}, \citenamefont {Xu}, \citenamefont {Zhang}, \citenamefont {Xu}, \citenamefont {Bi},\ and\ \citenamefont {Luo}}]{zhang2025PRB}%
  \BibitemOpen
  \bibfield  {author} {\bibinfo {author} {\bibfnamefont {C.}~\bibnamefont {Zhang}}, \bibinfo {author} {\bibfnamefont {Y.}~\bibnamefont {Xu}}, \bibinfo {author} {\bibfnamefont {J.-H.}\ \bibnamefont {Zhang}}, \bibinfo {author} {\bibfnamefont {C.}~\bibnamefont {Xu}}, \bibinfo {author} {\bibfnamefont {Z.}~\bibnamefont {Bi}},\ and\ \bibinfo {author} {\bibfnamefont {Z.-X.}\ \bibnamefont {Luo}},\ }\bibfield  {title} {\bibinfo {title} {{Strong-to-weak spontaneous breaking of 1-form symmetry and intrinsically mixed topological order}},\ }\href {https://doi.org/10.1103/PhysRevB.111.115137} {\bibfield  {journal} {\bibinfo  {journal} {Phys. Rev. B}\ }\textbf {\bibinfo {volume} {111}},\ \bibinfo {pages} {115137} (\bibinfo {year} {2025})}\BibitemShut {NoStop}%
\bibitem [{\citenamefont {Liu}\ \emph {et~al.}(2025)\citenamefont {Liu}, \citenamefont {Chen}, \citenamefont {Zhang}, \citenamefont {Zhou},\ and\ \citenamefont {Zhang}}]{liu2025CommsPhys}%
  \BibitemOpen
  \bibfield  {author} {\bibinfo {author} {\bibfnamefont {Z.}~\bibnamefont {Liu}}, \bibinfo {author} {\bibfnamefont {L.}~\bibnamefont {Chen}}, \bibinfo {author} {\bibfnamefont {Y.}~\bibnamefont {Zhang}}, \bibinfo {author} {\bibfnamefont {S.}~\bibnamefont {Zhou}},\ and\ \bibinfo {author} {\bibfnamefont {P.}~\bibnamefont {Zhang}},\ }\bibfield  {title} {\bibinfo {title} {{Diagnosing strong-to-weak symmetry breaking via Wightman correlators}},\ }\href {https://doi.org/https://doi.org/10.1038/s42005-025-02199-7} {\bibfield  {journal} {\bibinfo  {journal} {Commun. Phys.}\ }\textbf {\bibinfo {volume} {8}},\ \bibinfo {pages} {274} (\bibinfo {year} {2025})}\BibitemShut {NoStop}%
\bibitem [{\citenamefont {Weinstein}(2025)}]{weinstein2025PRL}%
  \BibitemOpen
  \bibfield  {author} {\bibinfo {author} {\bibfnamefont {Z.}~\bibnamefont {Weinstein}},\ }\bibfield  {title} {\bibinfo {title} {{Efficient Detection of Strong-to-Weak Spontaneous Symmetry Breaking via the R\'enyi-1 Correlator}},\ }\href {https://doi.org/10.1103/PhysRevLett.134.150405} {\bibfield  {journal} {\bibinfo  {journal} {Phys. Rev. Lett.}\ }\textbf {\bibinfo {volume} {134}},\ \bibinfo {pages} {150405} (\bibinfo {year} {2025})}\BibitemShut {NoStop}%
\bibitem [{\citenamefont {Huang}\ \emph {et~al.}(2025)\citenamefont {Huang}, \citenamefont {Qi}, \citenamefont {Zhang},\ and\ \citenamefont {Lucas}}]{huang2025PRB}%
  \BibitemOpen
  \bibfield  {author} {\bibinfo {author} {\bibfnamefont {X.}~\bibnamefont {Huang}}, \bibinfo {author} {\bibfnamefont {M.}~\bibnamefont {Qi}}, \bibinfo {author} {\bibfnamefont {J.-H.}\ \bibnamefont {Zhang}},\ and\ \bibinfo {author} {\bibfnamefont {A.}~\bibnamefont {Lucas}},\ }\bibfield  {title} {\bibinfo {title} {{Hydrodynamics as the effective field theory of strong-to-weak spontaneous symmetry breaking}},\ }\href {https://doi.org/10.1103/PhysRevB.111.125147} {\bibfield  {journal} {\bibinfo  {journal} {Phys. Rev. B}\ }\textbf {\bibinfo {volume} {111}},\ \bibinfo {pages} {125147} (\bibinfo {year} {2025})}\BibitemShut {NoStop}%
\bibitem [{\citenamefont {Ma}\ \emph {et~al.}(2025{\natexlab{a}})\citenamefont {Ma}, \citenamefont {Zhang}, \citenamefont {Bi}, \citenamefont {Cheng},\ and\ \citenamefont {Wang}}]{ma2025PRX}%
  \BibitemOpen
  \bibfield  {author} {\bibinfo {author} {\bibfnamefont {R.}~\bibnamefont {Ma}}, \bibinfo {author} {\bibfnamefont {J.-H.}\ \bibnamefont {Zhang}}, \bibinfo {author} {\bibfnamefont {Z.}~\bibnamefont {Bi}}, \bibinfo {author} {\bibfnamefont {M.}~\bibnamefont {Cheng}},\ and\ \bibinfo {author} {\bibfnamefont {C.}~\bibnamefont {Wang}},\ }\bibfield  {title} {\bibinfo {title} {{Topological Phases with Average Symmetries: The Decohered, the Disordered, and the Intrinsic}},\ }\href {https://doi.org/10.1103/PhysRevX.15.021062} {\bibfield  {journal} {\bibinfo  {journal} {Phys. Rev. X}\ }\textbf {\bibinfo {volume} {15}},\ \bibinfo {pages} {021062} (\bibinfo {year} {2025}{\natexlab{a}})}\BibitemShut {NoStop}%
\bibitem [{\citenamefont {Ma}\ and\ \citenamefont {Turzillo}(2025)}]{ma2025PRXQ}%
  \BibitemOpen
  \bibfield  {author} {\bibinfo {author} {\bibfnamefont {R.}~\bibnamefont {Ma}}\ and\ \bibinfo {author} {\bibfnamefont {A.}~\bibnamefont {Turzillo}},\ }\bibfield  {title} {\bibinfo {title} {{Symmetry-Protected Topological Phases of Mixed States in the Doubled Space}},\ }\href {https://doi.org/10.1103/PRXQuantum.6.010348} {\bibfield  {journal} {\bibinfo  {journal} {PRX Quantum}\ }\textbf {\bibinfo {volume} {6}},\ \bibinfo {pages} {010348} (\bibinfo {year} {2025})}\BibitemShut {NoStop}%
\bibitem [{\citenamefont {Guo}\ and\ \citenamefont {Yang}(2025)}]{guo2025PRB}%
  \BibitemOpen
  \bibfield  {author} {\bibinfo {author} {\bibfnamefont {Y.}~\bibnamefont {Guo}}\ and\ \bibinfo {author} {\bibfnamefont {S.}~\bibnamefont {Yang}},\ }\bibfield  {title} {\bibinfo {title} {{Strong-to-weak spontaneous symmetry breaking meets average symmetry-protected topological order}},\ }\href {https://doi.org/10.1103/PhysRevB.111.L201108} {\bibfield  {journal} {\bibinfo  {journal} {Phys. Rev. B}\ }\textbf {\bibinfo {volume} {111}},\ \bibinfo {pages} {L201108} (\bibinfo {year} {2025})}\BibitemShut {NoStop}%
\bibitem [{\citenamefont {Feng}\ \emph {et~al.}(2025)\citenamefont {Feng}, \citenamefont {Cheng},\ and\ \citenamefont {Ippoliti}}]{feng2025arXiv}%
  \BibitemOpen
  \bibfield  {author} {\bibinfo {author} {\bibfnamefont {X.}~\bibnamefont {Feng}}, \bibinfo {author} {\bibfnamefont {Z.}~\bibnamefont {Cheng}},\ and\ \bibinfo {author} {\bibfnamefont {M.}~\bibnamefont {Ippoliti}},\ }\bibfield  {title} {\bibinfo {title} {{Hardness of observing strong-to-weak symmetry breaking}},\ }\href {https://arxiv.org/abs/2504.12233} {\bibfield  {journal} {\bibinfo  {journal} {arXiv:2504.12233}\ } (\bibinfo {year} {2025})}\BibitemShut {NoStop}%
\bibitem [{\citenamefont {S\'a}\ and\ \citenamefont {B\'eri}(2025)}]{sa2025SWSSB}%
  \BibitemOpen
  \bibfield  {author} {\bibinfo {author} {\bibfnamefont {L.}~\bibnamefont {S\'a}}\ and\ \bibinfo {author} {\bibfnamefont {B.}~\bibnamefont {B\'eri}},\ }\bibfield  {title} {\bibinfo {title} {{Exactly solvable dissipative dynamics and one-form strong-to-weak spontaneous symmetry breaking in interacting two-dimensional spin systems}},\ }\href {https://doi.org/10.1103/1lzb-vp1w} {\bibfield  {journal} {\bibinfo  {journal} {Phys. Rev. B}\ }\textbf {\bibinfo {volume} {112}},\ \bibinfo {pages} {144311} (\bibinfo {year} {2025})}\BibitemShut {NoStop}%
\bibitem [{\citenamefont {Ziereis}\ \emph {et~al.}(2025)\citenamefont {Ziereis}, \citenamefont {Moudgalya},\ and\ \citenamefont {Knap}}]{ziereis2025arXiv}%
  \BibitemOpen
  \bibfield  {author} {\bibinfo {author} {\bibfnamefont {N.}~\bibnamefont {Ziereis}}, \bibinfo {author} {\bibfnamefont {S.}~\bibnamefont {Moudgalya}},\ and\ \bibinfo {author} {\bibfnamefont {M.}~\bibnamefont {Knap}},\ }\bibfield  {title} {\bibinfo {title} {{Strong-to-Weak Symmetry Breaking Phases in Steady States of Quantum Operations}},\ }\href {https://arxiv.org/abs/2509.09669} {\bibfield  {journal} {\bibinfo  {journal} {arXiv:2509.09669}\ } (\bibinfo {year} {2025})}\BibitemShut {NoStop}%
\bibitem [{\citenamefont {Li}\ \emph {et~al.}(2018)\citenamefont {Li}, \citenamefont {Chen},\ and\ \citenamefont {Fisher}}]{Li_2018_mipt}%
  \BibitemOpen
  \bibfield  {author} {\bibinfo {author} {\bibfnamefont {Y.}~\bibnamefont {Li}}, \bibinfo {author} {\bibfnamefont {X.}~\bibnamefont {Chen}},\ and\ \bibinfo {author} {\bibfnamefont {M.~P.~A.}\ \bibnamefont {Fisher}},\ }\bibfield  {title} {\bibinfo {title} {{Quantum Zeno effect and the many-body entanglement transition}},\ }\href {https://doi.org/10.1103/PhysRevB.98.205136} {\bibfield  {journal} {\bibinfo  {journal} {Phys. Rev. B}\ }\textbf {\bibinfo {volume} {98}},\ \bibinfo {pages} {205136} (\bibinfo {year} {2018})}\BibitemShut {NoStop}%
\bibitem [{\citenamefont {Skinner}\ \emph {et~al.}(2019)\citenamefont {Skinner}, \citenamefont {Ruhman},\ and\ \citenamefont {Nahum}}]{Skinner_2019_mipt}%
  \BibitemOpen
  \bibfield  {author} {\bibinfo {author} {\bibfnamefont {B.}~\bibnamefont {Skinner}}, \bibinfo {author} {\bibfnamefont {J.}~\bibnamefont {Ruhman}},\ and\ \bibinfo {author} {\bibfnamefont {A.}~\bibnamefont {Nahum}},\ }\bibfield  {title} {\bibinfo {title} {{Measurement-Induced Phase Transitions in the Dynamics of Entanglement}},\ }\href {https://doi.org/10.1103/PhysRevX.9.031009} {\bibfield  {journal} {\bibinfo  {journal} {Phys. Rev. X}\ }\textbf {\bibinfo {volume} {9}},\ \bibinfo {pages} {031009} (\bibinfo {year} {2019})}\BibitemShut {NoStop}%
\bibitem [{\citenamefont {Chan}\ \emph {et~al.}(2019)\citenamefont {Chan}, \citenamefont {Nandkishore}, \citenamefont {Pretko},\ and\ \citenamefont {Smith}}]{chan2019mipt}%
  \BibitemOpen
  \bibfield  {author} {\bibinfo {author} {\bibfnamefont {A.}~\bibnamefont {Chan}}, \bibinfo {author} {\bibfnamefont {R.~M.}\ \bibnamefont {Nandkishore}}, \bibinfo {author} {\bibfnamefont {M.}~\bibnamefont {Pretko}},\ and\ \bibinfo {author} {\bibfnamefont {G.}~\bibnamefont {Smith}},\ }\bibfield  {title} {\bibinfo {title} {{Unitary-projective entanglement dynamics}},\ }\href {https://doi.org/10.1103/PhysRevB.99.224307} {\bibfield  {journal} {\bibinfo  {journal} {Phys. Rev. B}\ }\textbf {\bibinfo {volume} {99}},\ \bibinfo {pages} {224307} (\bibinfo {year} {2019})}\BibitemShut {NoStop}%
\bibitem [{\citenamefont {Li}\ \emph {et~al.}(2019)\citenamefont {Li}, \citenamefont {Chen},\ and\ \citenamefont {Fisher}}]{Li_2019_mipt}%
  \BibitemOpen
  \bibfield  {author} {\bibinfo {author} {\bibfnamefont {Y.}~\bibnamefont {Li}}, \bibinfo {author} {\bibfnamefont {X.}~\bibnamefont {Chen}},\ and\ \bibinfo {author} {\bibfnamefont {M.~P.~A.}\ \bibnamefont {Fisher}},\ }\bibfield  {title} {\bibinfo {title} {{Measurement-driven entanglement transition in hybrid quantum circuits}},\ }\href {https://doi.org/10.1103/PhysRevB.100.134306} {\bibfield  {journal} {\bibinfo  {journal} {Phys. Rev. B}\ }\textbf {\bibinfo {volume} {100}},\ \bibinfo {pages} {134306} (\bibinfo {year} {2019})}\BibitemShut {NoStop}%
\bibitem [{\citenamefont {Szyniszewski}\ \emph {et~al.}(2019)\citenamefont {Szyniszewski}, \citenamefont {Romito},\ and\ \citenamefont {Schomerus}}]{Szyniszewski2019_mipt_circuits}%
  \BibitemOpen
  \bibfield  {author} {\bibinfo {author} {\bibfnamefont {M.}~\bibnamefont {Szyniszewski}}, \bibinfo {author} {\bibfnamefont {A.}~\bibnamefont {Romito}},\ and\ \bibinfo {author} {\bibfnamefont {H.}~\bibnamefont {Schomerus}},\ }\bibfield  {title} {\bibinfo {title} {{Entanglement transition from variable-strength weak measurements}},\ }\href {https://doi.org/10.1103/PhysRevB.100.064204} {\bibfield  {journal} {\bibinfo  {journal} {Phys. Rev. B}\ }\textbf {\bibinfo {volume} {100}},\ \bibinfo {pages} {064204} (\bibinfo {year} {2019})}\BibitemShut {NoStop}%
\bibitem [{\citenamefont {Jian}\ \emph {et~al.}(2020)\citenamefont {Jian}, \citenamefont {You}, \citenamefont {Vasseur},\ and\ \citenamefont {Ludwig}}]{jian2020PRB}%
  \BibitemOpen
  \bibfield  {author} {\bibinfo {author} {\bibfnamefont {C.-M.}\ \bibnamefont {Jian}}, \bibinfo {author} {\bibfnamefont {Y.-Z.}\ \bibnamefont {You}}, \bibinfo {author} {\bibfnamefont {R.}~\bibnamefont {Vasseur}},\ and\ \bibinfo {author} {\bibfnamefont {A.~W.~W.}\ \bibnamefont {Ludwig}},\ }\bibfield  {title} {\bibinfo {title} {{Measurement-induced criticality in random quantum circuits}},\ }\href {https://doi.org/10.1103/PhysRevB.101.104302} {\bibfield  {journal} {\bibinfo  {journal} {Phys. Rev. B}\ }\textbf {\bibinfo {volume} {101}},\ \bibinfo {pages} {104302} (\bibinfo {year} {2020})}\BibitemShut {NoStop}%
\bibitem [{\citenamefont {Bao}\ \emph {et~al.}(2020)\citenamefont {Bao}, \citenamefont {Choi},\ and\ \citenamefont {Altman}}]{bao2020PRB}%
  \BibitemOpen
  \bibfield  {author} {\bibinfo {author} {\bibfnamefont {Y.}~\bibnamefont {Bao}}, \bibinfo {author} {\bibfnamefont {S.}~\bibnamefont {Choi}},\ and\ \bibinfo {author} {\bibfnamefont {E.}~\bibnamefont {Altman}},\ }\bibfield  {title} {\bibinfo {title} {{Theory of the phase transition in random unitary circuits with measurements}},\ }\href {https://doi.org/10.1103/PhysRevB.101.104301} {\bibfield  {journal} {\bibinfo  {journal} {Phys. Rev. B}\ }\textbf {\bibinfo {volume} {101}},\ \bibinfo {pages} {104301} (\bibinfo {year} {2020})}\BibitemShut {NoStop}%
\bibitem [{\citenamefont {Choi}\ \emph {et~al.}(2020)\citenamefont {Choi}, \citenamefont {Bao}, \citenamefont {Qi},\ and\ \citenamefont {Altman}}]{choi2020PRL}%
  \BibitemOpen
  \bibfield  {author} {\bibinfo {author} {\bibfnamefont {S.}~\bibnamefont {Choi}}, \bibinfo {author} {\bibfnamefont {Y.}~\bibnamefont {Bao}}, \bibinfo {author} {\bibfnamefont {X.-L.}\ \bibnamefont {Qi}},\ and\ \bibinfo {author} {\bibfnamefont {E.}~\bibnamefont {Altman}},\ }\bibfield  {title} {\bibinfo {title} {{Quantum Error Correction in Scrambling Dynamics and Measurement-Induced Phase Transition}},\ }\href {https://doi.org/10.1103/PhysRevLett.125.030505} {\bibfield  {journal} {\bibinfo  {journal} {Phys. Rev. Lett.}\ }\textbf {\bibinfo {volume} {125}},\ \bibinfo {pages} {030505} (\bibinfo {year} {2020})}\BibitemShut {NoStop}%
\bibitem [{\citenamefont {Gullans}\ and\ \citenamefont {Huse}(2020)}]{gullans2020_mipt}%
  \BibitemOpen
  \bibfield  {author} {\bibinfo {author} {\bibfnamefont {M.~J.}\ \bibnamefont {Gullans}}\ and\ \bibinfo {author} {\bibfnamefont {D.~A.}\ \bibnamefont {Huse}},\ }\bibfield  {title} {\bibinfo {title} {{Dynamical Purification Phase Transition Induced by Quantum Measurements}},\ }\href {https://doi.org/10.1103/PhysRevX.10.041020} {\bibfield  {journal} {\bibinfo  {journal} {Phys. Rev. X}\ }\textbf {\bibinfo {volume} {10}},\ \bibinfo {pages} {041020} (\bibinfo {year} {2020})}\BibitemShut {NoStop}%
\bibitem [{\citenamefont {Turkeshi}\ \emph {et~al.}(2020)\citenamefont {Turkeshi}, \citenamefont {Fazio},\ and\ \citenamefont {Dalmonte}}]{Turkeshi2020_mipt_circuits}%
  \BibitemOpen
  \bibfield  {author} {\bibinfo {author} {\bibfnamefont {X.}~\bibnamefont {Turkeshi}}, \bibinfo {author} {\bibfnamefont {R.}~\bibnamefont {Fazio}},\ and\ \bibinfo {author} {\bibfnamefont {M.}~\bibnamefont {Dalmonte}},\ }\bibfield  {title} {\bibinfo {title} {{Measurement-induced criticality in $(2+1)$-dimensional hybrid quantum circuits}},\ }\href {https://doi.org/10.1103/PhysRevB.102.014315} {\bibfield  {journal} {\bibinfo  {journal} {Phys. Rev. B}\ }\textbf {\bibinfo {volume} {102}},\ \bibinfo {pages} {014315} (\bibinfo {year} {2020})}\BibitemShut {NoStop}%
\bibitem [{\citenamefont {Nahum}\ \emph {et~al.}(2021)\citenamefont {Nahum}, \citenamefont {Roy}, \citenamefont {Skinner},\ and\ \citenamefont {Ruhman}}]{Nahum2021_mipt_circuits}%
  \BibitemOpen
  \bibfield  {author} {\bibinfo {author} {\bibfnamefont {A.}~\bibnamefont {Nahum}}, \bibinfo {author} {\bibfnamefont {S.}~\bibnamefont {Roy}}, \bibinfo {author} {\bibfnamefont {B.}~\bibnamefont {Skinner}},\ and\ \bibinfo {author} {\bibfnamefont {J.}~\bibnamefont {Ruhman}},\ }\bibfield  {title} {\bibinfo {title} {{Measurement and Entanglement Phase Transitions in All-To-All Quantum Circuits, on Quantum Trees, and in Landau-Ginsburg Theory}},\ }\href {https://doi.org/10.1103/PRXQuantum.2.010352} {\bibfield  {journal} {\bibinfo  {journal} {PRX Quantum}\ }\textbf {\bibinfo {volume} {2}},\ \bibinfo {pages} {010352} (\bibinfo {year} {2021})}\BibitemShut {NoStop}%
\bibitem [{\citenamefont {Ippoliti}\ \emph {et~al.}(2021)\citenamefont {Ippoliti}, \citenamefont {Gullans}, \citenamefont {Gopalakrishnan}, \citenamefont {Huse},\ and\ \citenamefont {Khemani}}]{Ippoliti2021_ept_measurement_only_circuits}%
  \BibitemOpen
  \bibfield  {author} {\bibinfo {author} {\bibfnamefont {M.}~\bibnamefont {Ippoliti}}, \bibinfo {author} {\bibfnamefont {M.~J.}\ \bibnamefont {Gullans}}, \bibinfo {author} {\bibfnamefont {S.}~\bibnamefont {Gopalakrishnan}}, \bibinfo {author} {\bibfnamefont {D.~A.}\ \bibnamefont {Huse}},\ and\ \bibinfo {author} {\bibfnamefont {V.}~\bibnamefont {Khemani}},\ }\bibfield  {title} {\bibinfo {title} {{Entanglement Phase Transitions in Measurement-Only Dynamics}},\ }\href {https://doi.org/10.1103/PhysRevX.11.011030} {\bibfield  {journal} {\bibinfo  {journal} {Phys. Rev. X}\ }\textbf {\bibinfo {volume} {11}},\ \bibinfo {pages} {011030} (\bibinfo {year} {2021})}\BibitemShut {NoStop}%
\bibitem [{\citenamefont {Bao}\ \emph {et~al.}(2021)\citenamefont {Bao}, \citenamefont {Choi},\ and\ \citenamefont {Altman}}]{bao2021_circuits}%
  \BibitemOpen
  \bibfield  {author} {\bibinfo {author} {\bibfnamefont {Y.}~\bibnamefont {Bao}}, \bibinfo {author} {\bibfnamefont {S.}~\bibnamefont {Choi}},\ and\ \bibinfo {author} {\bibfnamefont {E.}~\bibnamefont {Altman}},\ }\bibfield  {title} {\bibinfo {title} {{Symmetry enriched phases of quantum circuits}},\ }\href {https://doi.org/https://doi.org/10.1016/j.aop.2021.168618} {\bibfield  {journal} {\bibinfo  {journal} {Ann. Phys.}\ }\textbf {\bibinfo {volume} {435}},\ \bibinfo {pages} {168618} (\bibinfo {year} {2021})}\BibitemShut {NoStop}%
\bibitem [{\citenamefont {Potter}(2022)}]{Potter2022review}%
  \BibitemOpen
  \bibfield  {author} {\bibinfo {author} {\bibfnamefont {R.}~\bibnamefont {Potter}, \bibfnamefont {Andrew C.and~Vasseur}},\ }\bibinfo {title} {E{ntanglement Dynamics in Hybrid Quantum Circuits}},\ in\ \href {https://doi.org/10.1007/978-3-031-03998-0_9} {\emph {\bibinfo {booktitle} {{Entanglement in Spin Chains: From Theory to Quantum Technology Applications}}}},\ \bibinfo {editor} {edited by\ \bibinfo {editor} {\bibfnamefont {A.}~\bibnamefont {Bayat}}, \bibinfo {editor} {\bibfnamefont {S.}~\bibnamefont {Bose}},\ and\ \bibinfo {editor} {\bibfnamefont {H.}~\bibnamefont {Johannesson}}}\ (\bibinfo  {publisher} {Springer},\ \bibinfo {address} {Cham},\ \bibinfo {year} {2022})\ pp.\ \bibinfo {pages} {211--249}\BibitemShut {NoStop}%
\bibitem [{\citenamefont {Fisher}\ \emph {et~al.}(2023)\citenamefont {Fisher}, \citenamefont {Khemani}, \citenamefont {Nahum},\ and\ \citenamefont {Vijay}}]{fisher2023review}%
  \BibitemOpen
  \bibfield  {author} {\bibinfo {author} {\bibfnamefont {M.~P.}\ \bibnamefont {Fisher}}, \bibinfo {author} {\bibfnamefont {V.}~\bibnamefont {Khemani}}, \bibinfo {author} {\bibfnamefont {A.}~\bibnamefont {Nahum}},\ and\ \bibinfo {author} {\bibfnamefont {S.}~\bibnamefont {Vijay}},\ }\bibfield  {title} {\bibinfo {title} {{Random Quantum Circuits}},\ }\href {https://doi.org/https://doi.org/10.1146/annurev-conmatphys-031720-030658} {\bibfield  {journal} {\bibinfo  {journal} {Annu. Rev. Condens. Matter Phys.}\ }\textbf {\bibinfo {volume} {14}},\ \bibinfo {pages} {335} (\bibinfo {year} {2023})}\BibitemShut {NoStop}%
\bibitem [{\citenamefont {Cao}\ \emph {et~al.}(2019)\citenamefont {Cao}, \citenamefont {Tilloy},\ and\ \citenamefont {Luca}}]{Cao_deLuca2019_entanglement_fermion_chain}%
  \BibitemOpen
  \bibfield  {author} {\bibinfo {author} {\bibfnamefont {X.}~\bibnamefont {Cao}}, \bibinfo {author} {\bibfnamefont {A.}~\bibnamefont {Tilloy}},\ and\ \bibinfo {author} {\bibfnamefont {A.~D.}\ \bibnamefont {Luca}},\ }\bibfield  {title} {\bibinfo {title} {{Entanglement in a fermion chain under continuous monitoring}},\ }\href {https://doi.org/10.21468/SciPostPhys.7.2.024} {\bibfield  {journal} {\bibinfo  {journal} {SciPost Phys.}\ }\textbf {\bibinfo {volume} {7}},\ \bibinfo {pages} {024} (\bibinfo {year} {2019})}\BibitemShut {NoStop}%
\bibitem [{\citenamefont {Alberton}\ \emph {et~al.}(2021)\citenamefont {Alberton}, \citenamefont {Buchhold},\ and\ \citenamefont {Diehl}}]{Alberton2021_entanglement_transition_free_fermion_chain}%
  \BibitemOpen
  \bibfield  {author} {\bibinfo {author} {\bibfnamefont {O.}~\bibnamefont {Alberton}}, \bibinfo {author} {\bibfnamefont {M.}~\bibnamefont {Buchhold}},\ and\ \bibinfo {author} {\bibfnamefont {S.}~\bibnamefont {Diehl}},\ }\bibfield  {title} {\bibinfo {title} {{Entanglement Transition in a Monitored Free-Fermion Chain: From Extended Criticality to Area Law}},\ }\href {https://doi.org/10.1103/PhysRevLett.126.170602} {\bibfield  {journal} {\bibinfo  {journal} {Phys. Rev. Lett.}\ }\textbf {\bibinfo {volume} {126}},\ \bibinfo {pages} {170602} (\bibinfo {year} {2021})}\BibitemShut {NoStop}%
\bibitem [{\citenamefont {Buchhold}\ \emph {et~al.}(2021)\citenamefont {Buchhold}, \citenamefont {Minoguchi}, \citenamefont {Altland},\ and\ \citenamefont {Diehl}}]{Buchhold2021_mipt}%
  \BibitemOpen
  \bibfield  {author} {\bibinfo {author} {\bibfnamefont {M.}~\bibnamefont {Buchhold}}, \bibinfo {author} {\bibfnamefont {Y.}~\bibnamefont {Minoguchi}}, \bibinfo {author} {\bibfnamefont {A.}~\bibnamefont {Altland}},\ and\ \bibinfo {author} {\bibfnamefont {S.}~\bibnamefont {Diehl}},\ }\bibfield  {title} {\bibinfo {title} {{Effective Theory for the Measurement-Induced Phase Transition of Dirac Fermions}},\ }\href {https://doi.org/10.1103/PhysRevX.11.041004} {\bibfield  {journal} {\bibinfo  {journal} {Phys. Rev. X}\ }\textbf {\bibinfo {volume} {11}},\ \bibinfo {pages} {041004} (\bibinfo {year} {2021})}\BibitemShut {NoStop}%
\bibitem [{\citenamefont {Fava}\ \emph {et~al.}(2023)\citenamefont {Fava}, \citenamefont {Piroli}, \citenamefont {Swann}, \citenamefont {Bernard},\ and\ \citenamefont {Nahum}}]{fava2023PRX}%
  \BibitemOpen
  \bibfield  {author} {\bibinfo {author} {\bibfnamefont {M.}~\bibnamefont {Fava}}, \bibinfo {author} {\bibfnamefont {L.}~\bibnamefont {Piroli}}, \bibinfo {author} {\bibfnamefont {T.}~\bibnamefont {Swann}}, \bibinfo {author} {\bibfnamefont {D.}~\bibnamefont {Bernard}},\ and\ \bibinfo {author} {\bibfnamefont {A.}~\bibnamefont {Nahum}},\ }\bibfield  {title} {\bibinfo {title} {{Nonlinear Sigma Models for Monitored Dynamics of Free Fermions}},\ }\href {https://doi.org/10.1103/PhysRevX.13.041045} {\bibfield  {journal} {\bibinfo  {journal} {Phys. Rev. X}\ }\textbf {\bibinfo {volume} {13}},\ \bibinfo {pages} {041045} (\bibinfo {year} {2023})}\BibitemShut {NoStop}%
\bibitem [{\citenamefont {Poboiko}\ \emph {et~al.}(2023)\citenamefont {Poboiko}, \citenamefont {P\"opperl}, \citenamefont {Gornyi},\ and\ \citenamefont {Mirlin}}]{Poboiko2023_free_fermions_no_mipt}%
  \BibitemOpen
  \bibfield  {author} {\bibinfo {author} {\bibfnamefont {I.}~\bibnamefont {Poboiko}}, \bibinfo {author} {\bibfnamefont {P.}~\bibnamefont {P\"opperl}}, \bibinfo {author} {\bibfnamefont {I.~V.}\ \bibnamefont {Gornyi}},\ and\ \bibinfo {author} {\bibfnamefont {A.~D.}\ \bibnamefont {Mirlin}},\ }\bibfield  {title} {\bibinfo {title} {{Theory of Free Fermions under Random Projective Measurements}},\ }\href {https://doi.org/10.1103/PhysRevX.13.041046} {\bibfield  {journal} {\bibinfo  {journal} {Phys. Rev. X}\ }\textbf {\bibinfo {volume} {13}},\ \bibinfo {pages} {041046} (\bibinfo {year} {2023})}\BibitemShut {NoStop}%
\bibitem [{\citenamefont {Poboiko}\ \emph {et~al.}(2024)\citenamefont {Poboiko}, \citenamefont {Gornyi},\ and\ \citenamefont {Mirlin}}]{Poboiko2024_mipt_free_fermion}%
  \BibitemOpen
  \bibfield  {author} {\bibinfo {author} {\bibfnamefont {I.}~\bibnamefont {Poboiko}}, \bibinfo {author} {\bibfnamefont {I.~V.}\ \bibnamefont {Gornyi}},\ and\ \bibinfo {author} {\bibfnamefont {A.~D.}\ \bibnamefont {Mirlin}},\ }\bibfield  {title} {\bibinfo {title} {{Measurement-Induced Phase Transition for Free Fermions above One Dimension}},\ }\href {https://doi.org/10.1103/PhysRevLett.132.110403} {\bibfield  {journal} {\bibinfo  {journal} {Phys. Rev. Lett.}\ }\textbf {\bibinfo {volume} {132}},\ \bibinfo {pages} {110403} (\bibinfo {year} {2024})}\BibitemShut {NoStop}%
\bibitem [{\citenamefont {Fava}\ \emph {et~al.}(2024)\citenamefont {Fava}, \citenamefont {Piroli}, \citenamefont {Bernard},\ and\ \citenamefont {Nahum}}]{Fava2024_mipt_free_fermion}%
  \BibitemOpen
  \bibfield  {author} {\bibinfo {author} {\bibfnamefont {M.}~\bibnamefont {Fava}}, \bibinfo {author} {\bibfnamefont {L.}~\bibnamefont {Piroli}}, \bibinfo {author} {\bibfnamefont {D.}~\bibnamefont {Bernard}},\ and\ \bibinfo {author} {\bibfnamefont {A.}~\bibnamefont {Nahum}},\ }\bibfield  {title} {\bibinfo {title} {{Monitored fermions with conserved $U(1)$ charge}},\ }\href {https://doi.org/10.1103/PhysRevResearch.6.043246} {\bibfield  {journal} {\bibinfo  {journal} {Phys. Rev. Res.}\ }\textbf {\bibinfo {volume} {6}},\ \bibinfo {pages} {043246} (\bibinfo {year} {2024})}\BibitemShut {NoStop}%
\bibitem [{\citenamefont {Carisch}\ \emph {et~al.}(2023)\citenamefont {Carisch}, \citenamefont {Romito},\ and\ \citenamefont {Zilberberg}}]{Carisch2023_mipt}%
  \BibitemOpen
  \bibfield  {author} {\bibinfo {author} {\bibfnamefont {C.}~\bibnamefont {Carisch}}, \bibinfo {author} {\bibfnamefont {A.}~\bibnamefont {Romito}},\ and\ \bibinfo {author} {\bibfnamefont {O.}~\bibnamefont {Zilberberg}},\ }\bibfield  {title} {\bibinfo {title} {{Quantifying measurement-induced quantum-to-classical crossover using an open-system entanglement measure}},\ }\href {https://doi.org/10.1103/PhysRevResearch.5.L042031} {\bibfield  {journal} {\bibinfo  {journal} {Phys. Rev. Res.}\ }\textbf {\bibinfo {volume} {5}},\ \bibinfo {pages} {L042031} (\bibinfo {year} {2023})}\BibitemShut {NoStop}%
\bibitem [{\citenamefont {Chahine}\ and\ \citenamefont {Buchhold}(2024)}]{chahine2024PRB}%
  \BibitemOpen
  \bibfield  {author} {\bibinfo {author} {\bibfnamefont {K.}~\bibnamefont {Chahine}}\ and\ \bibinfo {author} {\bibfnamefont {M.}~\bibnamefont {Buchhold}},\ }\bibfield  {title} {\bibinfo {title} {{Entanglement phases, localization, and multifractality of monitored free fermions in two dimensions}},\ }\href {https://doi.org/10.1103/PhysRevB.110.054313} {\bibfield  {journal} {\bibinfo  {journal} {Phys. Rev. B}\ }\textbf {\bibinfo {volume} {110}},\ \bibinfo {pages} {054313} (\bibinfo {year} {2024})}\BibitemShut {NoStop}%
\bibitem [{\citenamefont {Fan}\ \emph {et~al.}(2025{\natexlab{a}})\citenamefont {Fan}, \citenamefont {Yin},\ and\ \citenamefont {Garc{\'\i}a-Garc{\'\i}a}}]{fan2025}%
  \BibitemOpen
  \bibfield  {author} {\bibinfo {author} {\bibfnamefont {B.}~\bibnamefont {Fan}}, \bibinfo {author} {\bibfnamefont {C.}~\bibnamefont {Yin}},\ and\ \bibinfo {author} {\bibfnamefont {A.~M.}\ \bibnamefont {Garc{\'\i}a-Garc{\'\i}a}},\ }\bibfield  {title} {\bibinfo {title} {{How does the entanglement entropy of a many-body quantum system change after a single measurement?}},\ }\href {https://arxiv.org/abs/2504.04071} {\bibfield  {journal} {\bibinfo  {journal} {arXiv:2504.04071}\ } (\bibinfo {year} {2025}{\natexlab{a}})}\BibitemShut {NoStop}%
\bibitem [{\citenamefont {Fan}\ \emph {et~al.}(2025{\natexlab{b}})\citenamefont {Fan}, \citenamefont {Yin},\ and\ \citenamefont {Garc{\'\i}a-Garc{\'\i}a}}]{fan2025B}%
  \BibitemOpen
  \bibfield  {author} {\bibinfo {author} {\bibfnamefont {B.}~\bibnamefont {Fan}}, \bibinfo {author} {\bibfnamefont {C.}~\bibnamefont {Yin}},\ and\ \bibinfo {author} {\bibfnamefont {A.~M.}\ \bibnamefont {Garc{\'\i}a-Garc{\'\i}a}},\ }\bibfield  {title} {\bibinfo {title} {{Entanglement dynamics of monitored non-interacting fermions on Graphic-Processing-Units}},\ }\href {https://arxiv.org/abs/2508.18468} {\bibfield  {journal} {\bibinfo  {journal} {arXiv:2508.18468}\ } (\bibinfo {year} {2025}{\natexlab{b}})}\BibitemShut {NoStop}%
\bibitem [{\citenamefont {Eissler}\ \emph {et~al.}(2025)\citenamefont {Eissler}, \citenamefont {Lesanovsky},\ and\ \citenamefont {Carollo}}]{eisller2025PRA}%
  \BibitemOpen
  \bibfield  {author} {\bibinfo {author} {\bibfnamefont {M.}~\bibnamefont {Eissler}}, \bibinfo {author} {\bibfnamefont {I.}~\bibnamefont {Lesanovsky}},\ and\ \bibinfo {author} {\bibfnamefont {F.}~\bibnamefont {Carollo}},\ }\bibfield  {title} {\bibinfo {title} {{Unraveling-induced entanglement phase transition in diffusive trajectories of continuously monitored noninteracting fermionic systems}},\ }\href {https://doi.org/10.1103/PhysRevA.111.022205} {\bibfield  {journal} {\bibinfo  {journal} {Phys. Rev. A}\ }\textbf {\bibinfo {volume} {111}},\ \bibinfo {pages} {022205} (\bibinfo {year} {2025})}\BibitemShut {NoStop}%
\bibitem [{\citenamefont {Soares}\ \emph {et~al.}(2025)\citenamefont {Soares}, \citenamefont {Le~Gal},\ and\ \citenamefont {Schir\`o}}]{Soares2025_entanglement_transistion_fermion_chain}%
  \BibitemOpen
  \bibfield  {author} {\bibinfo {author} {\bibfnamefont {R.~D.}\ \bibnamefont {Soares}}, \bibinfo {author} {\bibfnamefont {Y.}~\bibnamefont {Le~Gal}},\ and\ \bibinfo {author} {\bibfnamefont {M.}~\bibnamefont {Schir\`o}},\ }\bibfield  {title} {\bibinfo {title} {{Entanglement transition due to particle losses in a monitored fermionic chain}},\ }\href {https://doi.org/10.1103/PhysRevB.111.064313} {\bibfield  {journal} {\bibinfo  {journal} {Phys. Rev. B}\ }\textbf {\bibinfo {volume} {111}},\ \bibinfo {pages} {064313} (\bibinfo {year} {2025})}\BibitemShut {NoStop}%
\bibitem [{\citenamefont {Ferrari}\ \emph {et~al.}(2025{\natexlab{a}})\citenamefont {Ferrari}, \citenamefont {Gravina}, \citenamefont {Eeltink}, \citenamefont {Scarlino}, \citenamefont {Savona},\ and\ \citenamefont {Minganti}}]{ferrari_2023_steadystatequantumchaosopen}%
  \BibitemOpen
  \bibfield  {author} {\bibinfo {author} {\bibfnamefont {F.}~\bibnamefont {Ferrari}}, \bibinfo {author} {\bibfnamefont {L.}~\bibnamefont {Gravina}}, \bibinfo {author} {\bibfnamefont {D.}~\bibnamefont {Eeltink}}, \bibinfo {author} {\bibfnamefont {P.}~\bibnamefont {Scarlino}}, \bibinfo {author} {\bibfnamefont {V.}~\bibnamefont {Savona}},\ and\ \bibinfo {author} {\bibfnamefont {F.}~\bibnamefont {Minganti}},\ }\bibfield  {title} {\bibinfo {title} {{Dissipative quantum chaos unveiled by stochastic quantum trajectories}},\ }\href {https://doi.org/10.1103/PhysRevResearch.7.013276} {\bibfield  {journal} {\bibinfo  {journal} {Phys. Rev. Res.}\ }\textbf {\bibinfo {volume} {7}},\ \bibinfo {pages} {013276} (\bibinfo {year} {2025}{\natexlab{a}})}\BibitemShut {NoStop}%
\bibitem [{\citenamefont {Prosen}(2012{\natexlab{a}})}]{prosen_2012_PT}%
  \BibitemOpen
  \bibfield  {author} {\bibinfo {author} {\bibfnamefont {T.}~\bibnamefont {Prosen}},\ }\bibfield  {title} {\bibinfo {title} {{$\mathbb{P}\mathbb{T}$-Symmetric Quantum Liouvillean Dynamics}},\ }\href {https://doi.org/10.1103/PhysRevLett.109.090404} {\bibfield  {journal} {\bibinfo  {journal} {Phys. Rev. Lett.}\ }\textbf {\bibinfo {volume} {109}},\ \bibinfo {pages} {090404} (\bibinfo {year} {2012}{\natexlab{a}})}\BibitemShut {NoStop}%
\bibitem [{\citenamefont {Prosen}(2012{\natexlab{b}})}]{prosen_2012_PT_2}%
  \BibitemOpen
  \bibfield  {author} {\bibinfo {author} {\bibfnamefont {T.}~\bibnamefont {Prosen}},\ }\bibfield  {title} {\bibinfo {title} {{Generic examples of $\mathbb{P}\mathbb{T}$-symmetric qubit (spin-1/2) Liouvillian dynamics}},\ }\href {https://doi.org/10.1103/PhysRevA.86.044103} {\bibfield  {journal} {\bibinfo  {journal} {Phys. Rev. A}\ }\textbf {\bibinfo {volume} {86}},\ \bibinfo {pages} {044103} (\bibinfo {year} {2012}{\natexlab{b}})}\BibitemShut {NoStop}%
\bibitem [{\citenamefont {Haga}(2023)}]{haga2023_non_steady_state_transition}%
  \BibitemOpen
  \bibfield  {author} {\bibinfo {author} {\bibfnamefont {T.}~\bibnamefont {Haga}},\ }\bibfield  {title} {\bibinfo {title} {{Spontaneous symmetry breaking in nonsteady modes of open quantum many-body systems}},\ }\href {https://doi.org/10.1103/PhysRevA.107.052208} {\bibfield  {journal} {\bibinfo  {journal} {Phys. Rev. A}\ }\textbf {\bibinfo {volume} {107}},\ \bibinfo {pages} {052208} (\bibinfo {year} {2023})}\BibitemShut {NoStop}%
\bibitem [{\citenamefont {Ma}\ \emph {et~al.}(2025{\natexlab{b}})\citenamefont {Ma}, \citenamefont {Guo}, \citenamefont {Gao}, \citenamefont {Papi\'{c}},\ and\ \citenamefont {Ying}}]{ma2025PT}%
  \BibitemOpen
  \bibfield  {author} {\bibinfo {author} {\bibfnamefont {J.-L.}\ \bibnamefont {Ma}}, \bibinfo {author} {\bibfnamefont {Z.}~\bibnamefont {Guo}}, \bibinfo {author} {\bibfnamefont {Y.}~\bibnamefont {Gao}}, \bibinfo {author} {\bibfnamefont {Z.}~\bibnamefont {Papi\'{c}}},\ and\ \bibinfo {author} {\bibfnamefont {L.}~\bibnamefont {Ying}},\ }\bibfield  {title} {\bibinfo {title} {{Liouvillian Spectral Transition in Noisy Quantum Many-Body Scars}},\ }\href {https://doi.org/10.1103/4my3-vk6c} {\bibfield  {journal} {\bibinfo  {journal} {Phys. Rev. Lett.}\ }\textbf {\bibinfo {volume} {135}},\ \bibinfo {pages} {180401} (\bibinfo {year} {2025}{\natexlab{b}})}\BibitemShut {NoStop}%
\bibitem [{\citenamefont {Wang}\ \emph {et~al.}(2020)\citenamefont {Wang}, \citenamefont {Piazza},\ and\ \citenamefont {Luitz}}]{wang2020_hierarchy}%
  \BibitemOpen
  \bibfield  {author} {\bibinfo {author} {\bibfnamefont {K.}~\bibnamefont {Wang}}, \bibinfo {author} {\bibfnamefont {F.}~\bibnamefont {Piazza}},\ and\ \bibinfo {author} {\bibfnamefont {D.~J.}\ \bibnamefont {Luitz}},\ }\bibfield  {title} {\bibinfo {title} {{Hierarchy of Relaxation Timescales in Local Random Liouvillians}},\ }\href {https://doi.org/10.1103/PhysRevLett.124.100604} {\bibfield  {journal} {\bibinfo  {journal} {Phys. Rev. Lett.}\ }\textbf {\bibinfo {volume} {124}},\ \bibinfo {pages} {100604} (\bibinfo {year} {2020})}\BibitemShut {NoStop}%
\bibitem [{\citenamefont {Sommer}\ \emph {et~al.}(2021)\citenamefont {Sommer}, \citenamefont {Piazza},\ and\ \citenamefont {Luitz}}]{sommer2021_hierarchy}%
  \BibitemOpen
  \bibfield  {author} {\bibinfo {author} {\bibfnamefont {O.~E.}\ \bibnamefont {Sommer}}, \bibinfo {author} {\bibfnamefont {F.}~\bibnamefont {Piazza}},\ and\ \bibinfo {author} {\bibfnamefont {D.~J.}\ \bibnamefont {Luitz}},\ }\bibfield  {title} {\bibinfo {title} {{Many-body hierarchy of dissipative timescales in a quantum computer}},\ }\href {https://doi.org/10.1103/PhysRevResearch.3.023190} {\bibfield  {journal} {\bibinfo  {journal} {Phys. Rev. Res.}\ }\textbf {\bibinfo {volume} {3}},\ \bibinfo {pages} {023190} (\bibinfo {year} {2021})}\BibitemShut {NoStop}%
\bibitem [{\citenamefont {Zhou}\ \emph {et~al.}(2021)\citenamefont {Zhou}, \citenamefont {Mao},\ and\ \citenamefont {Zhai}}]{zhou2021_separation}%
  \BibitemOpen
  \bibfield  {author} {\bibinfo {author} {\bibfnamefont {Y.-N.}\ \bibnamefont {Zhou}}, \bibinfo {author} {\bibfnamefont {L.}~\bibnamefont {Mao}},\ and\ \bibinfo {author} {\bibfnamefont {H.}~\bibnamefont {Zhai}},\ }\bibfield  {title} {\bibinfo {title} {{R\'enyi entropy dynamics and Lindblad spectrum for open quantum systems}},\ }\href {https://doi.org/10.1103/PhysRevResearch.3.043060} {\bibfield  {journal} {\bibinfo  {journal} {Phys. Rev. Res.}\ }\textbf {\bibinfo {volume} {3}},\ \bibinfo {pages} {043060} (\bibinfo {year} {2021})}\BibitemShut {NoStop}%
\bibitem [{\citenamefont {Popkov}\ and\ \citenamefont {Presilla}(2021)}]{popkov2021_spectrum}%
  \BibitemOpen
  \bibfield  {author} {\bibinfo {author} {\bibfnamefont {V.}~\bibnamefont {Popkov}}\ and\ \bibinfo {author} {\bibfnamefont {C.}~\bibnamefont {Presilla}},\ }\bibfield  {title} {\bibinfo {title} {{Full Spectrum of the Liouvillian of Open Dissipative Quantum Systems in the Zeno Limit}},\ }\href {https://doi.org/10.1103/PhysRevLett.126.190402} {\bibfield  {journal} {\bibinfo  {journal} {Phys. Rev. Lett.}\ }\textbf {\bibinfo {volume} {126}},\ \bibinfo {pages} {190402} (\bibinfo {year} {2021})}\BibitemShut {NoStop}%
\bibitem [{\citenamefont {Li}\ \emph {et~al.}(2022)\citenamefont {Li}, \citenamefont {Rose}, \citenamefont {Garrahan},\ and\ \citenamefont {Luitz}}]{li2023_metastability}%
  \BibitemOpen
  \bibfield  {author} {\bibinfo {author} {\bibfnamefont {J.~L.}\ \bibnamefont {Li}}, \bibinfo {author} {\bibfnamefont {D.~C.}\ \bibnamefont {Rose}}, \bibinfo {author} {\bibfnamefont {J.~P.}\ \bibnamefont {Garrahan}},\ and\ \bibinfo {author} {\bibfnamefont {D.~J.}\ \bibnamefont {Luitz}},\ }\bibfield  {title} {\bibinfo {title} {{Random matrix theory for quantum and classical metastability in local Liouvillians}},\ }\href {https://doi.org/10.1103/PhysRevB.105.L180201} {\bibfield  {journal} {\bibinfo  {journal} {Phys. Rev. B}\ }\textbf {\bibinfo {volume} {105}},\ \bibinfo {pages} {L180201} (\bibinfo {year} {2022})}\BibitemShut {NoStop}%
\bibitem [{\citenamefont {Shackleton}\ and\ \citenamefont {Scheurer}(2024)}]{shackleton2024_spin_liquid}%
  \BibitemOpen
  \bibfield  {author} {\bibinfo {author} {\bibfnamefont {H.}~\bibnamefont {Shackleton}}\ and\ \bibinfo {author} {\bibfnamefont {M.~S.}\ \bibnamefont {Scheurer}},\ }\bibfield  {title} {\bibinfo {title} {{Exactly solvable dissipative spin liquid}},\ }\href {https://doi.org/10.1103/PhysRevB.109.085115} {\bibfield  {journal} {\bibinfo  {journal} {Phys. Rev. B}\ }\textbf {\bibinfo {volume} {109}},\ \bibinfo {pages} {085115} (\bibinfo {year} {2024})}\BibitemShut {NoStop}%
\bibitem [{\citenamefont {Garc\'{\i}a-Garc\'{\i}a}\ \emph {et~al.}(2025)\citenamefont {Garc\'{\i}a-Garc\'{\i}a}, \citenamefont {S\'a}, \citenamefont {Verbaarschot},\ and\ \citenamefont {Yin}}]{garcia2025_topology}%
  \BibitemOpen
  \bibfield  {author} {\bibinfo {author} {\bibfnamefont {A.~M.}\ \bibnamefont {Garc\'{\i}a-Garc\'{\i}a}}, \bibinfo {author} {\bibfnamefont {L.}~\bibnamefont {S\'a}}, \bibinfo {author} {\bibfnamefont {J.~J.~M.}\ \bibnamefont {Verbaarschot}},\ and\ \bibinfo {author} {\bibfnamefont {C.}~\bibnamefont {Yin}},\ }\bibfield  {title} {\bibinfo {title} {{Emergent topology in many-body dissipative quantum matter}},\ }\href {https://doi.org/10.1103/PhysRevB.111.035157} {\bibfield  {journal} {\bibinfo  {journal} {Phys. Rev. B}\ }\textbf {\bibinfo {volume} {111}},\ \bibinfo {pages} {035157} (\bibinfo {year} {2025})}\BibitemShut {NoStop}%
\bibitem [{\citenamefont {Sá}\ \emph {et~al.}(2023)\citenamefont {Sá}, \citenamefont {Ribeiro},\ and\ \citenamefont {Prosen}}]{Sa_2023_SymmetryClass}%
  \BibitemOpen
  \bibfield  {author} {\bibinfo {author} {\bibfnamefont {L.}~\bibnamefont {Sá}}, \bibinfo {author} {\bibfnamefont {P.}~\bibnamefont {Ribeiro}},\ and\ \bibinfo {author} {\bibfnamefont {T.}~\bibnamefont {Prosen}},\ }\bibfield  {title} {\bibinfo {title} {{Symmetry Classification of Many-Body Lindbladians: Tenfold Way and Beyond}},\ }\href {https://doi.org/10.1103/PhysRevX.13.031019} {\bibfield  {journal} {\bibinfo  {journal} {Phys. Rev. X}\ }\textbf {\bibinfo {volume} {13}},\ \bibinfo {pages} {031019} (\bibinfo {year} {2023})}\BibitemShut {NoStop}%
\bibitem [{\citenamefont {Kawabata}\ \emph {et~al.}(2023)\citenamefont {Kawabata}, \citenamefont {Kulkarni}, \citenamefont {Li}, \citenamefont {Numasawa},\ and\ \citenamefont {Ryu}}]{kawabata2023_PRXQ}%
  \BibitemOpen
  \bibfield  {author} {\bibinfo {author} {\bibfnamefont {K.}~\bibnamefont {Kawabata}}, \bibinfo {author} {\bibfnamefont {A.}~\bibnamefont {Kulkarni}}, \bibinfo {author} {\bibfnamefont {J.}~\bibnamefont {Li}}, \bibinfo {author} {\bibfnamefont {T.}~\bibnamefont {Numasawa}},\ and\ \bibinfo {author} {\bibfnamefont {S.}~\bibnamefont {Ryu}},\ }\bibfield  {title} {\bibinfo {title} {{Symmetry of Open Quantum Systems: Classification of Dissipative Quantum Chaos}},\ }\href {https://doi.org/10.1103/PRXQuantum.4.030328} {\bibfield  {journal} {\bibinfo  {journal} {PRX Quantum}\ }\textbf {\bibinfo {volume} {4}},\ \bibinfo {pages} {030328} (\bibinfo {year} {2023})}\BibitemShut {NoStop}%
\bibitem [{\citenamefont {Garc\'{\i}a-Garc\'{\i}a}\ \emph {et~al.}(2024)\citenamefont {Garc\'{\i}a-Garc\'{\i}a}, \citenamefont {S\'a}, \citenamefont {Verbaarschot},\ and\ \citenamefont {Yin}}]{garcia2024PT}%
  \BibitemOpen
  \bibfield  {author} {\bibinfo {author} {\bibfnamefont {A.~M.}\ \bibnamefont {Garc\'{\i}a-Garc\'{\i}a}}, \bibinfo {author} {\bibfnamefont {L.}~\bibnamefont {S\'a}}, \bibinfo {author} {\bibfnamefont {J.~J.~M.}\ \bibnamefont {Verbaarschot}},\ and\ \bibinfo {author} {\bibfnamefont {C.}~\bibnamefont {Yin}},\ }\bibfield  {title} {\bibinfo {title} {{Toward a classification of PT-symmetric quantum systems: From dissipative dynamics to topology and wormholes}},\ }\href {https://doi.org/10.1103/PhysRevD.109.105017} {\bibfield  {journal} {\bibinfo  {journal} {Phys. Rev. D}\ }\textbf {\bibinfo {volume} {109}},\ \bibinfo {pages} {105017} (\bibinfo {year} {2024})}\BibitemShut {NoStop}%
\bibitem [{\citenamefont {Hamazaki}\ \emph {et~al.}(2022)\citenamefont {Hamazaki}, \citenamefont {Nakagawa}, \citenamefont {Haga},\ and\ \citenamefont {Ueda}}]{hamazaki2022_lindbladianmanybodylocalization}%
  \BibitemOpen
  \bibfield  {author} {\bibinfo {author} {\bibfnamefont {R.}~\bibnamefont {Hamazaki}}, \bibinfo {author} {\bibfnamefont {M.}~\bibnamefont {Nakagawa}}, \bibinfo {author} {\bibfnamefont {T.}~\bibnamefont {Haga}},\ and\ \bibinfo {author} {\bibfnamefont {M.}~\bibnamefont {Ueda}},\ }\bibfield  {title} {\bibinfo {title} {{Lindbladian Many-Body Localization}},\ }\href {https://arxiv.org/abs/2206.02984} {\bibfield  {journal} {\bibinfo  {journal} {arXiv:2206.02984}\ } (\bibinfo {year} {2022})}\BibitemShut {NoStop}%
\bibitem [{\citenamefont {Zhou}\ \emph {et~al.}(2023)\citenamefont {Zhou}, \citenamefont {Yu}, \citenamefont {Wu}, \citenamefont {Li}, \citenamefont {Zhang}, \citenamefont {Li},\ and\ \citenamefont {Chen}}]{Zhou2023_Mpemba}%
  \BibitemOpen
  \bibfield  {author} {\bibinfo {author} {\bibfnamefont {Y.-L.}\ \bibnamefont {Zhou}}, \bibinfo {author} {\bibfnamefont {X.-D.}\ \bibnamefont {Yu}}, \bibinfo {author} {\bibfnamefont {C.-W.}\ \bibnamefont {Wu}}, \bibinfo {author} {\bibfnamefont {X.-Q.}\ \bibnamefont {Li}}, \bibinfo {author} {\bibfnamefont {J.}~\bibnamefont {Zhang}}, \bibinfo {author} {\bibfnamefont {W.}~\bibnamefont {Li}},\ and\ \bibinfo {author} {\bibfnamefont {P.-X.}\ \bibnamefont {Chen}},\ }\bibfield  {title} {\bibinfo {title} {{Accelerating relaxation through Liouvillian exceptional point}},\ }\href {https://doi.org/10.1103/PhysRevResearch.5.043036} {\bibfield  {journal} {\bibinfo  {journal} {Phys. Rev. Res.}\ }\textbf {\bibinfo {volume} {5}},\ \bibinfo {pages} {043036} (\bibinfo {year} {2023})}\BibitemShut {NoStop}%
\bibitem [{\citenamefont {Dutta}\ \emph {et~al.}(2025)\citenamefont {Dutta}, \citenamefont {Zhang},\ and\ \citenamefont {Haque}}]{DuttaZhangHaque_PRL2025_LimitCycles}%
  \BibitemOpen
  \bibfield  {author} {\bibinfo {author} {\bibfnamefont {S.}~\bibnamefont {Dutta}}, \bibinfo {author} {\bibfnamefont {S.}~\bibnamefont {Zhang}},\ and\ \bibinfo {author} {\bibfnamefont {M.}~\bibnamefont {Haque}},\ }\bibfield  {title} {\bibinfo {title} {{Quantum Origin of Limit Cycles, Fixed Points, and Critical Slowing Down}},\ }\href {https://doi.org/10.1103/PhysRevLett.134.050407} {\bibfield  {journal} {\bibinfo  {journal} {Phys. Rev. Lett.}\ }\textbf {\bibinfo {volume} {134}},\ \bibinfo {pages} {050407} (\bibinfo {year} {2025})}\BibitemShut {NoStop}%
\bibitem [{\citenamefont {Chirame}\ and\ \citenamefont {Burnell1}(2025)}]{chirame2025}%
  \BibitemOpen
  \bibfield  {author} {\bibinfo {author} {\bibfnamefont {S.}~\bibnamefont {Chirame}}\ and\ \bibinfo {author} {\bibfnamefont {F.~J.}\ \bibnamefont {Burnell1}},\ }\bibfield  {title} {\bibinfo {title} {{Open system dynamics in local Lindbladians with chaotic spectra}},\ }\href {https://arxiv.org/abs/2510.15193} {\bibfield  {journal} {\bibinfo  {journal} {arXiv:2510.15193}\ } (\bibinfo {year} {2025})}\BibitemShut {NoStop}%
\bibitem [{\citenamefont {Lu}\ and\ \citenamefont {Raz}(2017)}]{Lu_Raz_2017_Mpemba}%
  \BibitemOpen
  \bibfield  {author} {\bibinfo {author} {\bibfnamefont {Z.}~\bibnamefont {Lu}}\ and\ \bibinfo {author} {\bibfnamefont {O.}~\bibnamefont {Raz}},\ }\bibfield  {title} {\bibinfo {title} {{Nonequilibrium thermodynamics of the Markovian Mpemba effect and its inverse}},\ }\href {https://doi.org/10.1073/pnas.1701264114} {\bibfield  {journal} {\bibinfo  {journal} {Proc. Natl Acad. Sci.}\ }\textbf {\bibinfo {volume} {114}},\ \bibinfo {pages} {5083} (\bibinfo {year} {2017})}\BibitemShut {NoStop}%
\bibitem [{\citenamefont {Ares}\ \emph {et~al.}(2025)\citenamefont {Ares}, \citenamefont {Calabrese},\ and\ \citenamefont {Murciano}}]{Ares2025_Mpemba_review}%
  \BibitemOpen
  \bibfield  {author} {\bibinfo {author} {\bibfnamefont {F.}~\bibnamefont {Ares}}, \bibinfo {author} {\bibfnamefont {P.}~\bibnamefont {Calabrese}},\ and\ \bibinfo {author} {\bibfnamefont {S.}~\bibnamefont {Murciano}},\ }\bibfield  {title} {\bibinfo {title} {{The quantum Mpemba effects}},\ }\href {https://doi.org/10.1038/s42254-025-00838-0} {\bibfield  {journal} {\bibinfo  {journal} {Nat. Rev. Phys.}\ }\textbf {\bibinfo {volume} {7}},\ \bibinfo {pages} {451} (\bibinfo {year} {2025})}\BibitemShut {NoStop}%
\bibitem [{\citenamefont {Manikandan}(2021)}]{Manikandan2021_Mpemba}%
  \BibitemOpen
  \bibfield  {author} {\bibinfo {author} {\bibfnamefont {S.~K.}\ \bibnamefont {Manikandan}},\ }\bibfield  {title} {\bibinfo {title} {{Equidistant quenches in few-level quantum systems}},\ }\href {https://doi.org/10.1103/PhysRevResearch.3.043108} {\bibfield  {journal} {\bibinfo  {journal} {Phys. Rev. Res.}\ }\textbf {\bibinfo {volume} {3}},\ \bibinfo {pages} {043108} (\bibinfo {year} {2021})}\BibitemShut {NoStop}%
\bibitem [{\citenamefont {Kochsiek}\ \emph {et~al.}(2022)\citenamefont {Kochsiek}, \citenamefont {Carollo},\ and\ \citenamefont {Lesanovsky}}]{Kochsiek2022_Mpemba}%
  \BibitemOpen
  \bibfield  {author} {\bibinfo {author} {\bibfnamefont {S.}~\bibnamefont {Kochsiek}}, \bibinfo {author} {\bibfnamefont {F.}~\bibnamefont {Carollo}},\ and\ \bibinfo {author} {\bibfnamefont {I.}~\bibnamefont {Lesanovsky}},\ }\bibfield  {title} {\bibinfo {title} {{Accelerating the approach of dissipative quantum spin systems towards stationarity through global spin rotations}},\ }\href {https://doi.org/10.1103/PhysRevA.106.012207} {\bibfield  {journal} {\bibinfo  {journal} {Phys. Rev. A}\ }\textbf {\bibinfo {volume} {106}},\ \bibinfo {pages} {012207} (\bibinfo {year} {2022})}\BibitemShut {NoStop}%
\bibitem [{\citenamefont {Ivander}\ \emph {et~al.}(2023)\citenamefont {Ivander}, \citenamefont {Anto-Sztrikacs},\ and\ \citenamefont {Segal}}]{Ivander2023_Mpemba}%
  \BibitemOpen
  \bibfield  {author} {\bibinfo {author} {\bibfnamefont {F.}~\bibnamefont {Ivander}}, \bibinfo {author} {\bibfnamefont {N.}~\bibnamefont {Anto-Sztrikacs}},\ and\ \bibinfo {author} {\bibfnamefont {D.}~\bibnamefont {Segal}},\ }\bibfield  {title} {\bibinfo {title} {{Hyperacceleration of quantum thermalization dynamics by bypassing long-lived coherences: An analytical treatment}},\ }\href {https://doi.org/10.1103/PhysRevE.108.014130} {\bibfield  {journal} {\bibinfo  {journal} {Phys. Rev. E}\ }\textbf {\bibinfo {volume} {108}},\ \bibinfo {pages} {014130} (\bibinfo {year} {2023})}\BibitemShut {NoStop}%
\bibitem [{\citenamefont {Carollo}\ \emph {et~al.}(2021)\citenamefont {Carollo}, \citenamefont {Lasanta},\ and\ \citenamefont {Lesanovsky}}]{carollo_2021_mpemba}%
  \BibitemOpen
  \bibfield  {author} {\bibinfo {author} {\bibfnamefont {F.}~\bibnamefont {Carollo}}, \bibinfo {author} {\bibfnamefont {A.}~\bibnamefont {Lasanta}},\ and\ \bibinfo {author} {\bibfnamefont {I.}~\bibnamefont {Lesanovsky}},\ }\bibfield  {title} {\bibinfo {title} {{Exponentially Accelerated Approach to Stationarity in Markovian Open Quantum Systems through the Mpemba Effect}},\ }\href {https://doi.org/10.1103/PhysRevLett.127.060401} {\bibfield  {journal} {\bibinfo  {journal} {Phys. Rev. Lett.}\ }\textbf {\bibinfo {volume} {127}},\ \bibinfo {pages} {060401} (\bibinfo {year} {2021})}\BibitemShut {NoStop}%
\bibitem [{\citenamefont {Chatterjee}\ \emph {et~al.}(2023)\citenamefont {Chatterjee}, \citenamefont {Takada},\ and\ \citenamefont {Hayakawa}}]{chatterjee_2023_mpemba}%
  \BibitemOpen
  \bibfield  {author} {\bibinfo {author} {\bibfnamefont {A.~K.}\ \bibnamefont {Chatterjee}}, \bibinfo {author} {\bibfnamefont {S.}~\bibnamefont {Takada}},\ and\ \bibinfo {author} {\bibfnamefont {H.}~\bibnamefont {Hayakawa}},\ }\bibfield  {title} {\bibinfo {title} {{Quantum Mpemba Effect in a Quantum Dot with Reservoirs}},\ }\href {https://doi.org/10.1103/PhysRevLett.131.080402} {\bibfield  {journal} {\bibinfo  {journal} {Phys. Rev. Lett.}\ }\textbf {\bibinfo {volume} {131}},\ \bibinfo {pages} {080402} (\bibinfo {year} {2023})}\BibitemShut {NoStop}%
\bibitem [{\citenamefont {Nava}\ and\ \citenamefont {Egger}(2024)}]{Nava2024_Mpemba}%
  \BibitemOpen
  \bibfield  {author} {\bibinfo {author} {\bibfnamefont {A.}~\bibnamefont {Nava}}\ and\ \bibinfo {author} {\bibfnamefont {R.}~\bibnamefont {Egger}},\ }\bibfield  {title} {\bibinfo {title} {{Mpemba Effects in Open Nonequilibrium Quantum Systems}},\ }\href {https://doi.org/10.1103/PhysRevLett.133.136302} {\bibfield  {journal} {\bibinfo  {journal} {Phys. Rev. Lett.}\ }\textbf {\bibinfo {volume} {133}},\ \bibinfo {pages} {136302} (\bibinfo {year} {2024})}\BibitemShut {NoStop}%
\bibitem [{\citenamefont {Wang}\ and\ \citenamefont {Wang}(2024)}]{Wang2024_Mpemba}%
  \BibitemOpen
  \bibfield  {author} {\bibinfo {author} {\bibfnamefont {X.}~\bibnamefont {Wang}}\ and\ \bibinfo {author} {\bibfnamefont {J.}~\bibnamefont {Wang}},\ }\bibfield  {title} {\bibinfo {title} {{Mpemba effects in nonequilibrium open quantum systems}},\ }\href {https://doi.org/10.1103/PhysRevResearch.6.033330} {\bibfield  {journal} {\bibinfo  {journal} {Phys. Rev. Res.}\ }\textbf {\bibinfo {volume} {6}},\ \bibinfo {pages} {033330} (\bibinfo {year} {2024})}\BibitemShut {NoStop}%
\bibitem [{\citenamefont {Liu}\ \emph {et~al.}(2024)\citenamefont {Liu}, \citenamefont {Yuan}, \citenamefont {Ruan}, \citenamefont {Xu}, \citenamefont {Luo}, \citenamefont {He}, \citenamefont {He}, \citenamefont {Ma},\ and\ \citenamefont {Wang}}]{Liu2024_Mpemba}%
  \BibitemOpen
  \bibfield  {author} {\bibinfo {author} {\bibfnamefont {D.}~\bibnamefont {Liu}}, \bibinfo {author} {\bibfnamefont {J.}~\bibnamefont {Yuan}}, \bibinfo {author} {\bibfnamefont {H.}~\bibnamefont {Ruan}}, \bibinfo {author} {\bibfnamefont {Y.}~\bibnamefont {Xu}}, \bibinfo {author} {\bibfnamefont {S.}~\bibnamefont {Luo}}, \bibinfo {author} {\bibfnamefont {J.}~\bibnamefont {He}}, \bibinfo {author} {\bibfnamefont {X.}~\bibnamefont {He}}, \bibinfo {author} {\bibfnamefont {Y.}~\bibnamefont {Ma}},\ and\ \bibinfo {author} {\bibfnamefont {J.}~\bibnamefont {Wang}},\ }\bibfield  {title} {\bibinfo {title} {{Speeding up quantum heat engines by the Mpemba effect}},\ }\href {https://doi.org/10.1103/PhysRevA.110.042218} {\bibfield  {journal} {\bibinfo  {journal} {Phys. Rev. A}\ }\textbf {\bibinfo {volume} {110}},\ \bibinfo {pages} {042218} (\bibinfo {year} {2024})}\BibitemShut {NoStop}%
\bibitem [{\citenamefont {Graf}\ \emph {et~al.}(2025)\citenamefont {Graf}, \citenamefont {Splettstoesser},\ and\ \citenamefont {Monsel}}]{Graf2025_Mpemba}%
  \BibitemOpen
  \bibfield  {author} {\bibinfo {author} {\bibfnamefont {J.}~\bibnamefont {Graf}}, \bibinfo {author} {\bibfnamefont {J.}~\bibnamefont {Splettstoesser}},\ and\ \bibinfo {author} {\bibfnamefont {J.}~\bibnamefont {Monsel}},\ }\bibfield  {title} {\bibinfo {title} {{Role of electron–electron interaction in the Mpemba effect in quantum dots}},\ }\href {https://doi.org/10.1088/1361-648X/adc3e3} {\bibfield  {journal} {\bibinfo  {journal} {J. Phys.: Condens. Matter}\ }\textbf {\bibinfo {volume} {37}},\ \bibinfo {pages} {195302} (\bibinfo {year} {2025})}\BibitemShut {NoStop}%
\bibitem [{\citenamefont {Furtado}\ and\ \citenamefont {Santos}(2025)}]{Furtado2025_Mpemba}%
  \BibitemOpen
  \bibfield  {author} {\bibinfo {author} {\bibfnamefont {J.}~\bibnamefont {Furtado}}\ and\ \bibinfo {author} {\bibfnamefont {A.~C.}\ \bibnamefont {Santos}},\ }\bibfield  {title} {\bibinfo {title} {{Enhanced quantum Mpemba effect with squeezed thermal reservoirs}},\ }\href {https://doi.org/https://doi.org/10.1016/j.aop.2025.170135} {\bibfield  {journal} {\bibinfo  {journal} {Ann. Phys.}\ }\textbf {\bibinfo {volume} {480}},\ \bibinfo {pages} {170135} (\bibinfo {year} {2025})}\BibitemShut {NoStop}%
\bibitem [{\citenamefont {Longhi}(2025)}]{Longhi2025_mpembaeffectsuper}%
  \BibitemOpen
  \bibfield  {author} {\bibinfo {author} {\bibfnamefont {S.}~\bibnamefont {Longhi}},\ }\bibfield  {title} {\bibinfo {title} {{Mpemba effect and super-accelerated thermalization in the damped quantum harmonic oscillator}},\ }\href {https://doi.org/10.22331/q-2025-03-26-1677} {\bibfield  {journal} {\bibinfo  {journal} {{Quantum}}\ }\textbf {\bibinfo {volume} {9}},\ \bibinfo {pages} {1677} (\bibinfo {year} {2025})}\BibitemShut {NoStop}%
\bibitem [{\citenamefont {Cipolloni}\ and\ \citenamefont {Kudler-Flam}(2024)}]{cipolloni2024PRB}%
  \BibitemOpen
  \bibfield  {author} {\bibinfo {author} {\bibfnamefont {G.}~\bibnamefont {Cipolloni}}\ and\ \bibinfo {author} {\bibfnamefont {J.}~\bibnamefont {Kudler-Flam}},\ }\bibfield  {title} {\bibinfo {title} {{Non-Hermitian Hamiltonians violate the eigenstate thermalization hypothesis}},\ }\href {https://doi.org/10.1103/PhysRevB.109.L020201} {\bibfield  {journal} {\bibinfo  {journal} {Phys. Rev. B}\ }\textbf {\bibinfo {volume} {109}},\ \bibinfo {pages} {L020201} (\bibinfo {year} {2024})}\BibitemShut {NoStop}%
\bibitem [{\citenamefont {Singha~Roy}\ \emph {et~al.}(2025)\citenamefont {Singha~Roy}, \citenamefont {Bandyopadhyay}, \citenamefont {Costa~de Almeida},\ and\ \citenamefont {Hauke}}]{Hauke_PRL2025_nonHermitianETH}%
  \BibitemOpen
  \bibfield  {author} {\bibinfo {author} {\bibfnamefont {S.}~\bibnamefont {Singha~Roy}}, \bibinfo {author} {\bibfnamefont {S.}~\bibnamefont {Bandyopadhyay}}, \bibinfo {author} {\bibfnamefont {R.}~\bibnamefont {Costa~de Almeida}},\ and\ \bibinfo {author} {\bibfnamefont {P.}~\bibnamefont {Hauke}},\ }\bibfield  {title} {\bibinfo {title} {{Unveiling Eigenstate Thermalization for Non-Hermitian systems}},\ }\href {https://doi.org/10.1103/PhysRevLett.134.180405} {\bibfield  {journal} {\bibinfo  {journal} {Phys. Rev. Lett.}\ }\textbf {\bibinfo {volume} {134}},\ \bibinfo {pages} {180405} (\bibinfo {year} {2025})}\BibitemShut {NoStop}%
\bibitem [{\citenamefont {Almeida}\ \emph {et~al.}(2025)\citenamefont {Almeida}, \citenamefont {Ribeiro}, \citenamefont {Haque},\ and\ \citenamefont {Sá}}]{almeida2025_lindblad_eth}%
  \BibitemOpen
  \bibfield  {author} {\bibinfo {author} {\bibfnamefont {G.}~\bibnamefont {Almeida}}, \bibinfo {author} {\bibfnamefont {P.}~\bibnamefont {Ribeiro}}, \bibinfo {author} {\bibfnamefont {M.}~\bibnamefont {Haque}},\ and\ \bibinfo {author} {\bibfnamefont {L.}~\bibnamefont {Sá}},\ }\bibfield  {title} {\bibinfo {title} {{Universality, Robustness, and Limits of the Eigenstate Thermalization Hypothesis in Open Quantum Systems}},\ }\href {https://arxiv.org/abs/2504.10261} {\bibfield  {journal} {\bibinfo  {journal} {arXiv:2504.10261}\ } (\bibinfo {year} {2025})}\BibitemShut {NoStop}%
\bibitem [{\citenamefont {Ferrari}\ \emph {et~al.}(2025{\natexlab{b}})\citenamefont {Ferrari}, \citenamefont {Savona},\ and\ \citenamefont {Minganti}}]{ferrari2025eth}%
  \BibitemOpen
  \bibfield  {author} {\bibinfo {author} {\bibfnamefont {F.}~\bibnamefont {Ferrari}}, \bibinfo {author} {\bibfnamefont {V.}~\bibnamefont {Savona}},\ and\ \bibinfo {author} {\bibfnamefont {F.}~\bibnamefont {Minganti}},\ }\bibfield  {title} {\bibinfo {title} {{Chaos and thermalization in open quantum systems}},\ }\href {https://arxiv.org/abs/2505.18260} {\bibfield  {journal} {\bibinfo  {journal} {arXiv:2505.18260}\ } (\bibinfo {year} {2025}{\natexlab{b}})}\BibitemShut {NoStop}%
\bibitem [{\citenamefont {Lindblad}(1976)}]{Lindblad_1976}%
  \BibitemOpen
  \bibfield  {author} {\bibinfo {author} {\bibfnamefont {G.}~\bibnamefont {Lindblad}},\ }\bibfield  {title} {\bibinfo {title} {{On the generators of quantum dynamical semigroups}},\ }\href {https://doi.org/10.1007/BF01608499} {\bibfield  {journal} {\bibinfo  {journal} {Commun. Math. Phys.}\ }\textbf {\bibinfo {volume} {48}},\ \bibinfo {pages} {119} (\bibinfo {year} {1976})}\BibitemShut {NoStop}%
\bibitem [{\citenamefont {Gorini}\ \emph {et~al.}(1976)\citenamefont {Gorini}, \citenamefont {Kossakowski},\ and\ \citenamefont {Sudarshan}}]{GKS_1976}%
  \BibitemOpen
  \bibfield  {author} {\bibinfo {author} {\bibfnamefont {V.}~\bibnamefont {Gorini}}, \bibinfo {author} {\bibfnamefont {A.}~\bibnamefont {Kossakowski}},\ and\ \bibinfo {author} {\bibfnamefont {E.~C.~G.}\ \bibnamefont {Sudarshan}},\ }\bibfield  {title} {\bibinfo {title} {{Completely positive dynamical semigroups of N‐level systems}},\ }\href {https://doi.org/10.1063/1.522979} {\bibfield  {journal} {\bibinfo  {journal} {J. Math. Phys.}\ }\textbf {\bibinfo {volume} {17}},\ \bibinfo {pages} {821} (\bibinfo {year} {1976})}\BibitemShut {NoStop}%
\bibitem [{\citenamefont {Carmichael}(1993)}]{carmichael_open_system_approach}%
  \BibitemOpen
  \bibfield  {author} {\bibinfo {author} {\bibfnamefont {H.}~\bibnamefont {Carmichael}},\ }\href {https://doi.org/10.1007/978-3-540-47620-7} {\emph {\bibinfo {title} {{An Open Systems Approach to Quantum Optics}}}}\ (\bibinfo  {publisher} {Springer Berlin},\ \bibinfo {address} {Heidelberg},\ \bibinfo {year} {1993})\BibitemShut {NoStop}%
\bibitem [{\citenamefont {Dalibard}\ \emph {et~al.}(1992)\citenamefont {Dalibard}, \citenamefont {Castin},\ and\ \citenamefont {M\o{}lmer}}]{Dalibard1992_wavefu_approach}%
  \BibitemOpen
  \bibfield  {author} {\bibinfo {author} {\bibfnamefont {J.}~\bibnamefont {Dalibard}}, \bibinfo {author} {\bibfnamefont {Y.}~\bibnamefont {Castin}},\ and\ \bibinfo {author} {\bibfnamefont {K.}~\bibnamefont {M\o{}lmer}},\ }\bibfield  {title} {\bibinfo {title} {{Wave-function approach to dissipative processes in quantum optics}},\ }\href {https://doi.org/10.1103/PhysRevLett.68.580} {\bibfield  {journal} {\bibinfo  {journal} {Phys. Rev. Lett.}\ }\textbf {\bibinfo {volume} {68}},\ \bibinfo {pages} {580} (\bibinfo {year} {1992})}\BibitemShut {NoStop}%
\bibitem [{\citenamefont {M{\o}lmer}\ \emph {et~al.}(1993)\citenamefont {M{\o}lmer}, \citenamefont {Castin},\ and\ \citenamefont {Dalibard}}]{Molmer_93_mcwf}%
  \BibitemOpen
  \bibfield  {author} {\bibinfo {author} {\bibfnamefont {K.}~\bibnamefont {M{\o}lmer}}, \bibinfo {author} {\bibfnamefont {Y.}~\bibnamefont {Castin}},\ and\ \bibinfo {author} {\bibfnamefont {J.}~\bibnamefont {Dalibard}},\ }\bibfield  {title} {\bibinfo {title} {{Monte Carlo wave-function method in quantum optics}},\ }\href {https://doi.org/10.1364/JOSAB.10.000524} {\bibfield  {journal} {\bibinfo  {journal} {J. Opt. Soc. Am. B}\ }\textbf {\bibinfo {volume} {10}},\ \bibinfo {pages} {524} (\bibinfo {year} {1993})}\BibitemShut {NoStop}%
\bibitem [{\citenamefont {Dum}\ \emph {et~al.}(1992)\citenamefont {Dum}, \citenamefont {Zoller},\ and\ \citenamefont {Ritsch}}]{DumZollerRitsch1992_MC_atomic_master_eq}%
  \BibitemOpen
  \bibfield  {author} {\bibinfo {author} {\bibfnamefont {R.}~\bibnamefont {Dum}}, \bibinfo {author} {\bibfnamefont {P.}~\bibnamefont {Zoller}},\ and\ \bibinfo {author} {\bibfnamefont {H.}~\bibnamefont {Ritsch}},\ }\bibfield  {title} {\bibinfo {title} {{Monte Carlo simulation of the atomic master equation for spontaneous emission}},\ }\href {https://doi.org/10.1103/PhysRevA.45.4879} {\bibfield  {journal} {\bibinfo  {journal} {Phys. Rev. A}\ }\textbf {\bibinfo {volume} {45}},\ \bibinfo {pages} {4879} (\bibinfo {year} {1992})}\BibitemShut {NoStop}%
\bibitem [{\citenamefont {Gisin}\ and\ \citenamefont {Percival}(1992)}]{Gisin_1992}%
  \BibitemOpen
  \bibfield  {author} {\bibinfo {author} {\bibfnamefont {N.}~\bibnamefont {Gisin}}\ and\ \bibinfo {author} {\bibfnamefont {I.~C.}\ \bibnamefont {Percival}},\ }\bibfield  {title} {\bibinfo {title} {{The quantum-state diffusion model applied to open systems}},\ }\href {https://doi.org/10.1088/0305-4470/25/21/023} {\bibfield  {journal} {\bibinfo  {journal} {J. Phys. A: Math. Gen.}\ }\textbf {\bibinfo {volume} {25}},\ \bibinfo {pages} {5677} (\bibinfo {year} {1992})}\BibitemShut {NoStop}%
\bibitem [{\citenamefont {Plenio}\ and\ \citenamefont {Knight}(1998)}]{plenio1998_quantum_jump_approach}%
  \BibitemOpen
  \bibfield  {author} {\bibinfo {author} {\bibfnamefont {M.~B.}\ \bibnamefont {Plenio}}\ and\ \bibinfo {author} {\bibfnamefont {P.~L.}\ \bibnamefont {Knight}},\ }\bibfield  {title} {\bibinfo {title} {{The quantum-jump approach to dissipative dynamics in quantum optics}},\ }\href {https://doi.org/10.1103/RevModPhys.70.101} {\bibfield  {journal} {\bibinfo  {journal} {Rev. Mod. Phys.}\ }\textbf {\bibinfo {volume} {70}},\ \bibinfo {pages} {101} (\bibinfo {year} {1998})}\BibitemShut {NoStop}%
\bibitem [{\citenamefont {Daley}(2014)}]{Daley2014}%
  \BibitemOpen
  \bibfield  {author} {\bibinfo {author} {\bibfnamefont {A.~J.}\ \bibnamefont {Daley}},\ }\bibfield  {title} {\bibinfo {title} {{Quantum trajectories and open many-body quantum systems}},\ }\href {https://doi.org/10.1080/00018732.2014.933502} {\bibfield  {journal} {\bibinfo  {journal} {Adv. Phys.}\ }\textbf {\bibinfo {volume} {63}},\ \bibinfo {pages} {77} (\bibinfo {year} {2014})}\BibitemShut {NoStop}%
\bibitem [{\citenamefont {Wiseman}\ and\ \citenamefont {Milburn}(1993)}]{wiseman1993PRA}%
  \BibitemOpen
  \bibfield  {author} {\bibinfo {author} {\bibfnamefont {H.~M.}\ \bibnamefont {Wiseman}}\ and\ \bibinfo {author} {\bibfnamefont {G.~J.}\ \bibnamefont {Milburn}},\ }\bibfield  {title} {\bibinfo {title} {{Interpretation of quantum jump and diffusion processes illustrated on the Bloch sphere}},\ }\href {https://doi.org/10.1103/PhysRevA.47.1652} {\bibfield  {journal} {\bibinfo  {journal} {Phys. Rev. A}\ }\textbf {\bibinfo {volume} {47}},\ \bibinfo {pages} {1652} (\bibinfo {year} {1993})}\BibitemShut {NoStop}%
\bibitem [{\citenamefont {Gardiner}\ and\ \citenamefont {Zoller}(2004)}]{gardiner2004_quantum_noise}%
  \BibitemOpen
  \bibfield  {author} {\bibinfo {author} {\bibfnamefont {C.}~\bibnamefont {Gardiner}}\ and\ \bibinfo {author} {\bibfnamefont {P.}~\bibnamefont {Zoller}},\ }\href {https://books.google.de/books?id=a_xsT8oGhdgC} {\emph {\bibinfo {title} {{Quantum Noise}}}}\ (\bibinfo  {publisher} {Springer Berlin},\ \bibinfo {address} {Heidelberg},\ \bibinfo {year} {2004})\BibitemShut {NoStop}%
\bibitem [{\citenamefont {Wiseman}\ and\ \citenamefont {Milburn}(2009)}]{wiseman2009book}%
  \BibitemOpen
  \bibfield  {author} {\bibinfo {author} {\bibfnamefont {H.~M.}\ \bibnamefont {Wiseman}}\ and\ \bibinfo {author} {\bibfnamefont {G.~J.}\ \bibnamefont {Milburn}},\ }\href {https://doi.org/10.1017/CBO9780511813948} {\emph {\bibinfo {title} {{Quantum measurement and control}}}}\ (\bibinfo  {publisher} {Cambridge University Press},\ \bibinfo {address} {Cambridge},\ \bibinfo {year} {2009})\BibitemShut {NoStop}%
\bibitem [{\citenamefont {Turkeshi}\ \emph {et~al.}(2021)\citenamefont {Turkeshi}, \citenamefont {Biella}, \citenamefont {Fazio}, \citenamefont {Dalmonte},\ and\ \citenamefont {Schir\'o}}]{Turkeshi2021_mipt_ising}%
  \BibitemOpen
  \bibfield  {author} {\bibinfo {author} {\bibfnamefont {X.}~\bibnamefont {Turkeshi}}, \bibinfo {author} {\bibfnamefont {A.}~\bibnamefont {Biella}}, \bibinfo {author} {\bibfnamefont {R.}~\bibnamefont {Fazio}}, \bibinfo {author} {\bibfnamefont {M.}~\bibnamefont {Dalmonte}},\ and\ \bibinfo {author} {\bibfnamefont {M.}~\bibnamefont {Schir\'o}},\ }\bibfield  {title} {\bibinfo {title} {{Measurement-induced entanglement transitions in the quantum Ising chain: From infinite to zero clicks}},\ }\href {https://doi.org/10.1103/PhysRevB.103.224210} {\bibfield  {journal} {\bibinfo  {journal} {Phys. Rev. B}\ }\textbf {\bibinfo {volume} {103}},\ \bibinfo {pages} {224210} (\bibinfo {year} {2021})}\BibitemShut {NoStop}%
\bibitem [{\citenamefont {Turkeshi}\ \emph {et~al.}(2024)\citenamefont {Turkeshi}, \citenamefont {Piroli},\ and\ \citenamefont {Schir\`o}}]{Turkeshi2024_monitored_fermionic_chains}%
  \BibitemOpen
  \bibfield  {author} {\bibinfo {author} {\bibfnamefont {X.}~\bibnamefont {Turkeshi}}, \bibinfo {author} {\bibfnamefont {L.}~\bibnamefont {Piroli}},\ and\ \bibinfo {author} {\bibfnamefont {M.}~\bibnamefont {Schir\`o}},\ }\bibfield  {title} {\bibinfo {title} {{Density and current statistics in boundary-driven monitored fermionic chains}},\ }\href {https://doi.org/10.1103/PhysRevB.109.144306} {\bibfield  {journal} {\bibinfo  {journal} {Phys. Rev. B}\ }\textbf {\bibinfo {volume} {109}},\ \bibinfo {pages} {144306} (\bibinfo {year} {2024})}\BibitemShut {NoStop}%
\bibitem [{\citenamefont {Pi\~nol}\ \emph {et~al.}(2024)\citenamefont {Pi\~nol}, \citenamefont {Mavrogordatos}, \citenamefont {Keys}, \citenamefont {Veyron}, \citenamefont {Sierant}, \citenamefont {Angel Garc\'{\i}a-March}, \citenamefont {Grandi}, \citenamefont {Mitchell}, \citenamefont {Wehr},\ and\ \citenamefont {Lewenstein}}]{Pinol2024_different_unraveling}%
  \BibitemOpen
  \bibfield  {author} {\bibinfo {author} {\bibfnamefont {E.}~\bibnamefont {Pi\~nol}}, \bibinfo {author} {\bibfnamefont {T.~K.}\ \bibnamefont {Mavrogordatos}}, \bibinfo {author} {\bibfnamefont {D.}~\bibnamefont {Keys}}, \bibinfo {author} {\bibfnamefont {R.}~\bibnamefont {Veyron}}, \bibinfo {author} {\bibfnamefont {P.}~\bibnamefont {Sierant}}, \bibinfo {author} {\bibfnamefont {M.}~\bibnamefont {Angel Garc\'{\i}a-March}}, \bibinfo {author} {\bibfnamefont {S.}~\bibnamefont {Grandi}}, \bibinfo {author} {\bibfnamefont {M.~W.}\ \bibnamefont {Mitchell}}, \bibinfo {author} {\bibfnamefont {J.}~\bibnamefont {Wehr}},\ and\ \bibinfo {author} {\bibfnamefont {M.}~\bibnamefont {Lewenstein}},\ }\bibfield  {title} {\bibinfo {title} {{Telling different unravelings apart via nonlinear quantum-trajectory averages}},\ }\href {https://doi.org/10.1103/PhysRevResearch.6.L032057} {\bibfield  {journal} {\bibinfo  {journal} {Phys. Rev. Res.}\ }\textbf {\bibinfo {volume} {6}},\ \bibinfo {pages} {L032057} (\bibinfo {year}
  {2024})}\BibitemShut {NoStop}%
\bibitem [{\citenamefont {Piccitto}\ \emph {et~al.}(2024)\citenamefont {Piccitto}, \citenamefont {Rossini},\ and\ \citenamefont {Russomanno}}]{Piccitto_2024_free_fermion_different_unravellings}%
  \BibitemOpen
  \bibfield  {author} {\bibinfo {author} {\bibfnamefont {G.}~\bibnamefont {Piccitto}}, \bibinfo {author} {\bibfnamefont {D.}~\bibnamefont {Rossini}},\ and\ \bibinfo {author} {\bibfnamefont {A.}~\bibnamefont {Russomanno}},\ }\bibfield  {title} {\bibinfo {title} {{The impact of different unravelings in a monitored system of free fermions}},\ }\href {https://doi.org/10.1140/epjb/s10051-024-00725-0} {\bibfield  {journal} {\bibinfo  {journal} {Eur. Phys. J. B}\ }\textbf {\bibinfo {volume} {97}},\ \bibinfo {pages} {90} (\bibinfo {year} {2024})}\BibitemShut {NoStop}%
\bibitem [{\citenamefont {Prosen}(2011)}]{Prosen2011_exact_ness}%
  \BibitemOpen
  \bibfield  {author} {\bibinfo {author} {\bibfnamefont {T.}~\bibnamefont {Prosen}},\ }\bibfield  {title} {\bibinfo {title} {{Exact Nonequilibrium Steady State of a Strongly Driven Open XXZ Chain}},\ }\href {https://doi.org/10.1103/PhysRevLett.107.137201} {\bibfield  {journal} {\bibinfo  {journal} {Phys. Rev. Lett.}\ }\textbf {\bibinfo {volume} {107}},\ \bibinfo {pages} {137201} (\bibinfo {year} {2011})}\BibitemShut {NoStop}%
\bibitem [{\citenamefont {Prosen}(2012{\natexlab{c}})}]{Prosen2012_comments_on_XXZ_chain}%
  \BibitemOpen
  \bibfield  {author} {\bibinfo {author} {\bibfnamefont {T.}~\bibnamefont {Prosen}},\ }\bibfield  {title} {\bibinfo {title} {{Comments on a boundary-driven open XXZ chain: asymmetric driving and uniqueness of steady states}},\ }\href {https://doi.org/10.1088/0031-8949/86/05/058511} {\bibfield  {journal} {\bibinfo  {journal} {Phys. Scr.}\ }\textbf {\bibinfo {volume} {86}},\ \bibinfo {pages} {058511} (\bibinfo {year} {2012}{\natexlab{c}})}\BibitemShut {NoStop}%
\bibitem [{\citenamefont {Prosen}(2015)}]{Prosen2015_spin_chain_review}%
  \BibitemOpen
  \bibfield  {author} {\bibinfo {author} {\bibfnamefont {T.}~\bibnamefont {Prosen}},\ }\bibfield  {title} {\bibinfo {title} {{Matrix product solutions of boundary driven quantum chains}},\ }\href {https://doi.org/10.1088/1751-8113/48/37/373001} {\bibfield  {journal} {\bibinfo  {journal} {J. Phys. A: Math. Theor.}\ }\textbf {\bibinfo {volume} {48}},\ \bibinfo {pages} {373001} (\bibinfo {year} {2015})}\BibitemShut {NoStop}%
\bibitem [{\citenamefont {Chalker}\ and\ \citenamefont {Mehlig}(1998)}]{chalker_mehlig_1998_eigenvector_statistics}%
  \BibitemOpen
  \bibfield  {author} {\bibinfo {author} {\bibfnamefont {J.~T.}\ \bibnamefont {Chalker}}\ and\ \bibinfo {author} {\bibfnamefont {B.}~\bibnamefont {Mehlig}},\ }\bibfield  {title} {\bibinfo {title} {{Eigenvector Statistics in Non-Hermitian Random Matrix Ensembles}},\ }\href {https://doi.org/10.1103/PhysRevLett.81.3367} {\bibfield  {journal} {\bibinfo  {journal} {Phys. Rev. Lett.}\ }\textbf {\bibinfo {volume} {81}},\ \bibinfo {pages} {3367} (\bibinfo {year} {1998})}\BibitemShut {NoStop}%
\bibitem [{\citenamefont {Mehlig}\ and\ \citenamefont {Chalker}(2000)}]{mehlig_chalker_2000_eigenvector_properties}%
  \BibitemOpen
  \bibfield  {author} {\bibinfo {author} {\bibfnamefont {B.}~\bibnamefont {Mehlig}}\ and\ \bibinfo {author} {\bibfnamefont {J.~T.}\ \bibnamefont {Chalker}},\ }\bibfield  {title} {\bibinfo {title} {{Statistical properties of eigenvectors in non-Hermitian Gaussian random matrix ensembles}},\ }\href {https://doi.org/10.1063/1.533302} {\bibfield  {journal} {\bibinfo  {journal} {J. Math. Phys.}\ }\textbf {\bibinfo {volume} {41}},\ \bibinfo {pages} {3233} (\bibinfo {year} {2000})}\BibitemShut {NoStop}%
\bibitem [{\citenamefont {Diehl}\ \emph {et~al.}(2008)\citenamefont {Diehl}, \citenamefont {Micheli}, \citenamefont {Kantian}, \citenamefont {Kraus}, \citenamefont {B{\"u}chler},\ and\ \citenamefont {Zoller}}]{Diehl2008_pure_ss_cold_atoms}%
  \BibitemOpen
  \bibfield  {author} {\bibinfo {author} {\bibfnamefont {S.}~\bibnamefont {Diehl}}, \bibinfo {author} {\bibfnamefont {A.}~\bibnamefont {Micheli}}, \bibinfo {author} {\bibfnamefont {A.}~\bibnamefont {Kantian}}, \bibinfo {author} {\bibfnamefont {B.}~\bibnamefont {Kraus}}, \bibinfo {author} {\bibfnamefont {H.~P.}\ \bibnamefont {B{\"u}chler}},\ and\ \bibinfo {author} {\bibfnamefont {P.}~\bibnamefont {Zoller}},\ }\bibfield  {title} {\bibinfo {title} {{Quantum states and phases in driven open quantum systems with cold atoms}},\ }\href {https://doi.org/10.1038/nphys1073} {\bibfield  {journal} {\bibinfo  {journal} {Nat. Phys.}\ }\textbf {\bibinfo {volume} {4}},\ \bibinfo {pages} {878} (\bibinfo {year} {2008})}\BibitemShut {NoStop}%
\bibitem [{\citenamefont {Kraus}\ \emph {et~al.}(2008)\citenamefont {Kraus}, \citenamefont {B\"uchler}, \citenamefont {Diehl}, \citenamefont {Kantian}, \citenamefont {Micheli},\ and\ \citenamefont {Zoller}}]{Kraus2008_prep_entangled_ss}%
  \BibitemOpen
  \bibfield  {author} {\bibinfo {author} {\bibfnamefont {B.}~\bibnamefont {Kraus}}, \bibinfo {author} {\bibfnamefont {H.~P.}\ \bibnamefont {B\"uchler}}, \bibinfo {author} {\bibfnamefont {S.}~\bibnamefont {Diehl}}, \bibinfo {author} {\bibfnamefont {A.}~\bibnamefont {Kantian}}, \bibinfo {author} {\bibfnamefont {A.}~\bibnamefont {Micheli}},\ and\ \bibinfo {author} {\bibfnamefont {P.}~\bibnamefont {Zoller}},\ }\bibfield  {title} {\bibinfo {title} {{Preparation of entangled states by quantum Markov processes}},\ }\href {https://doi.org/10.1103/PhysRevA.78.042307} {\bibfield  {journal} {\bibinfo  {journal} {Phys. Rev. A}\ }\textbf {\bibinfo {volume} {78}},\ \bibinfo {pages} {042307} (\bibinfo {year} {2008})}\BibitemShut {NoStop}%
\end{thebibliography}%

\appendix{}

\section*{End Matter}

\section{Numerical Details}

\emph{Vectorization.}---%
Numerically, we represent $\mathcal{L}$ in Eq.~\eqref{eq:lindblad_me} as a linear operator acting on a doubled Hilbert space $\mathcal{H}\otimes\mathcal{H}^*$:
\begin{equation}
\label{eq:liouvillian_supop}
\begin{split}
    \mathcal{L} = &-i \left(H \otimes \mathds{1} - \mathds{1} \otimes H^*\right) 
    \\
    &+\sum_k \gamma_k \left(L_{k} \otimes L_{k}^* - \frac{1}{2}  L_k^\dagger L_k \otimes \mathds{1} -\frac{1}{2} \mathds{1} \otimes (L_k^\dagger L_k)^* \right).
\end{split}
\end{equation}
In this picture, the Liouvillian is represented by a $D^2\times D^2$ matrix and the density matrix $\rho$ by a $ D^2$-dimensional vector $|\rho\rangle\!\rangle$. The steady state is obtained directly via exact diagonalization of the matrix representation of $\mathcal{L}$.

\emph{Quantum jump protocol.}---%
To unravel the quantum master equation into quantum trajectories, we use the Monte Carlo wave function method \cite{carmichael_open_system_approach,Dalibard1992_wavefu_approach,Molmer_93_mcwf,DumZollerRitsch1992_MC_atomic_master_eq,plenio1998_quantum_jump_approach,Daley2014}. Therein, the time evolution of a quantum trajectory $|\psi(t)\rangle$ consists of a non-unitary deterministic evolution interrupted by jumps at random times.
The probability of such a jump to occur within a small time step $\delta t$ is given by
\begin{align}
    \delta p = \delta t \sum_k \gamma_k \langle \psi(t) | L_k^\dagger L_k |\psi(t) \rangle \equiv \sum_k \delta p_k.
\end{align}
In case there is a jump, the time-evolved state is given by one of the $k$ possible normalized states 
\begin{align}
    |\psi(t+\delta t)\rangle = \frac{1}{\sqrt{\delta p_k / \delta t}} L_k |\psi(t)\rangle,
\end{align}
with probability $\delta p_k / \delta p$.
With probability $1-\delta p$, no jump occurs, and we have deterministic evolution with the effective (non-Hermitian) Hamiltonian
\begin{align}
    H_\text{eff} = H - \frac{i}{2} \sum_k \gamma_k L_k^\dagger L_k,
\end{align}
such that, for a small time step $\delta t$, the time-evolved state (after renormalizing) becomes
\begin{align}
    |\psi(t+\delta t)\rangle = \frac{1}{\sqrt{1-\delta p}} (1 - i H_\text{eff} \delta t) |\psi(t) \rangle.
\end{align} 

\emph{Sampling late-time properties.}---%
To compute the CM and IPR, we need to sample quantum trajectories in the limit of $t\to\infty$. However, in practice, the stochastic propagation of each trajectory has to be terminated at a finite $t_\text{ss}$.
To track convergence towards $\rho_\text{ss}$, we monitor the running average of the CM over 20 and 50 time steps; we deem the trajectory to be sufficiently converged when the difference of the two averages is less than the root mean square of their standard deviations.

\emph{Degeneracies.}---%
In general, a Liouvillian can have degenerate eigenvalues; for example, this is the case in the XXZ model discussed in the Main Text. For a degenerate eigenvalue, the corresponding left and right eigenspaces are not one-dimensional, and the freedom in the choice of the basis of this space amounts to additional freedom in the definition of overlaps with the eigenstates. We lift this indeterminancy by enforcing Eq.~(\ref{eq:conjugate_coefficients}) as follows.

Let $\{r_\alpha\}_{\alpha=1}^n$ and $\{r_{\alpha'}\}_{\alpha'=1}^n$ be the set of right eigenvectors corresponding to an $n$-fold degenerate eigenvalue $\lambda$ and its complex conjugate $\lambda^*$, respectively (we again assume that the Liouvillian is diagonalizable).
For Eq.~\eqref{eq:conjugate_coefficients} to hold, we need to
ensure that $r_{\alpha'}=r_\alpha^\dagger$, which requires rotating the eigenbasis in the degenerate subspace of either $\lambda$ or $\lambda^*$.
In the particular case of a degenerate real eigenstate $\lambda = \lambda^*$, this means rotating the basis such that $r_\alpha = r_\alpha^\dagger$.
However, this still does not fix the products $p_\alpha = c_\alpha d_\alpha$ uniquely (although their sum over the degenerate subspace and thus the CM are fixed): We resolve the remaining ambiguity by assigning to all $n$ eigenstates of this subspace the same quasi probability $\tilde p_\alpha = \sum_\alpha p_\alpha / n$.

\section{Bosonic systems}

To exemplify the versatility of our results, we consider two bosonic models with very different properties, in addition to the XXZ chain discussed in the Main Text.

\emph{Driven Bose-Hubbard dimer.}---%
We start with the example of a Bose-Hubbard dimer subject to driving~\cite{ferrari_2023_steadystatequantumchaosopen}.
With bosonic creation and annihilation operators $a_j^\dagger, a_j$ at two sites $j=1,2$, the Hamiltonian is expressed as
\begin{align}
    H = &- J \left(a_2^\dagger a_1 + a_1^\dagger a_2\right) +\sum_{j=1,2} \left( -\Delta a_j^\dagger a_j + \frac{U}{2} a_j^{\dagger\,2} a_j^2 \right)
     \nonumber\\
        &+ F\left(a_1^\dagger + a_1\right), \label{eq:hamiltonian_bhdimer}
\end{align}
where the $\Delta$ is the detuning, $U$ the Kerr nonlinearity, $J$ the hopping strength, and $F$ the amplitude of the driving field \cite{ferrari_2023_steadystatequantumchaosopen}. We fix $U=1$ and $J=2$ and vary $\Delta\in[-4,10]$ and $F\in[1,4]$.
To implement loss at both sites, the jump operators are chosen as the annihilation operators,
\begin{align}
\label{eq:jump_bhdimer}
    L_{j} = a_{j}, \quad j = 1,2,
\end{align}
with uniform coupling $\gamma_1=\gamma_2 \equiv\gamma= 1$.
Due to the driving $F\neq0$, particle number is not conserved and $H$ is defined on an infinite-dimensional Hilbert space. Numerically, we choose a cavity cutoff $N_c=5$ for the particle number, letting the Hamiltonian act only in the Hilbert space spanned by the states $|n_1,n_2\rangle$ with $n_1 + n_2 \leq N_c$.
As the initial state for the quantum trajectories, we choose a coherent state $|\alpha\rangle|\alpha\rangle$ with $\alpha = 3 \sqrt{F/(D-i\gamma)}$.

\emph{Bose-Hubbard chain.}---%
For a chain of length $N$ with periodic boundary conditions, the Bose-Hubbard Hamiltonian is given by
\begin{align} 
    H = -J \sum_{j, j+1} (a_j^\dagger a_{j+1}+\mathrm{h.c.}) + \frac{U}{2} \sum_{j} a_j^{\dagger\,2} a_j^2.
    \label{eq:hamiltonian_bhchain}
\end{align}
We consider a chain with $N=4$ sites and parameters $J\in[-0.5,0.9]$ and $U\in[-2,10]$. To implement dissipation, we consider the jump operators
\begin{align}
    L_j = (a_j^\dagger + a_{j+1}^\dagger)(a_j - a_{j+1}),\quad j = 1, \dots, N,
    \label{eq:jumpoperators_bhchain}
\end{align}
acting on neighboring sites with uniform coupling strength $\gamma_j = \gamma = 1$. 
With these jump operators, for $U=0$, the steady state is pure and given by 
the Bose-Einstein condensate~\cite{Diehl2008_pure_ss_cold_atoms, Kraus2008_prep_entangled_ss}
\begin{align}
    |\mathrm{BEC}\rangle = \frac{(a_{q=0}^\dagger)^{N_b}}{\sqrt{{N_b}!}} |\mathrm{vac}\rangle,
\end{align}
where the momentum-space creation operators $a_q^\dagger = (1/N)\sum_j a_j^\dagger e^{-iqx_j}$ act on the vaccum state $|\mathrm{vac}\rangle$ that is annihilated by all the $a_j$.

The particle number operator, $N_b=\sum_{j=1}^N a_j^\dagger a_j$, commutes with the Hamiltonian and each jump operator. Therefore, it is a strong symmetry \cite{BucaProsen2012_note_on_symmetries} of the Liouvillian and it is conserved for each individual quantum trajectory. We fix the number of bosons to be $N_b=3$.
The initial state for the quantum trajectories is taken to be a state with random coefficients in the Fock representation, which remains fixed for all trajectories.

\begin{figure}[t]
\centering
\includegraphics[width=\linewidth]{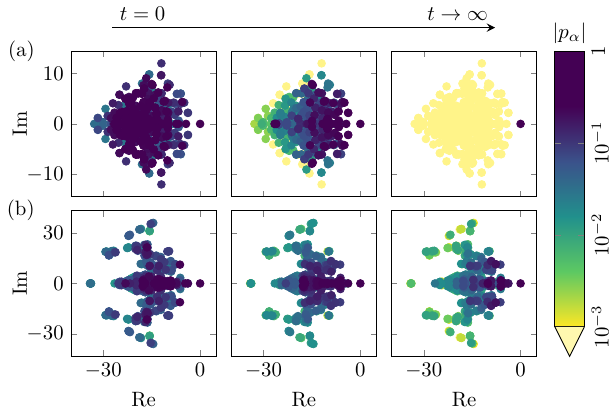}
\caption{
Same as Fig.~\ref{fig:spectra_evo} for the Bose-Hubbard chain, Eqs.~(\ref{eq:hamiltonian_bhchain}) and (\ref{eq:jumpoperators_bhchain}), with $N=4$, $N_b=3$,  $J=1$. (a) $U=0$, pure steady state. (b) $U=10$, mixed steady state. 
    }
    \label{fig:spectra_evo_bhc}
\end{figure}

\begin{figure}[t]
    \centering
    \includegraphics[width=\linewidth]{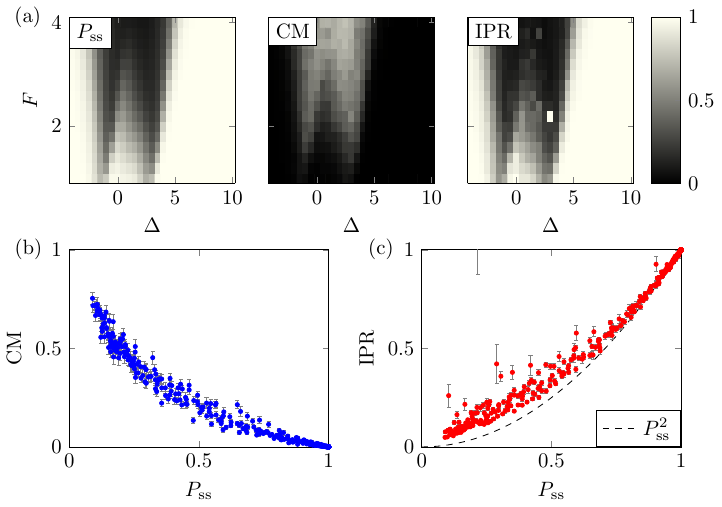}
    \caption{
    Same as Fig.~\ref{fig:indicators_xxz} for the Bose-Hubbard dimer with parameters $\Delta$ and $F$.
    }
    \label{fig:indicators_bhd}
\end{figure}

\begin{figure}[t]
    \centering
    \includegraphics[width=\linewidth]{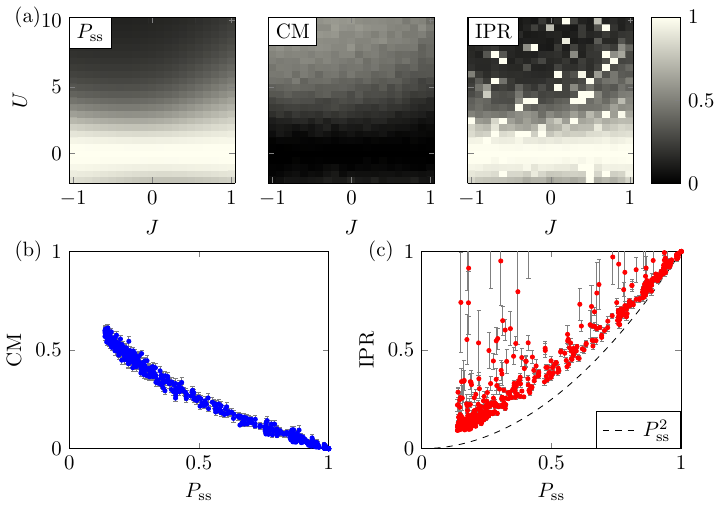}
    \caption{
    Same as Fig.~\ref{fig:indicators_xxz} for the Bose-Hubbard chain with parameters $J$ and $U$.
    }
    \label{fig:indicators_bhc}
\end{figure}

\emph{Numerics.}---%
In Figs.~\ref{fig:spectra_evo}(c) and (d) of the Main Text, we showed that depending on the choice of $\Delta$, quantum trajectories for the Bose-Hubbard dimer can either localize or delocalize in the Liouvillian eigenbasis. Similar results are found for the Bose-Hubbard chain, see Fig.~\ref{fig:spectra_evo_bhc}.
For the case of $U=0$, the steady state of the Bose-Hubbard chain is pure and consequently the overlaps of the trajectories with every other eigenstate decay to zero. This special case leads to the localization of the trajectories at late times, with $\mathrm{CM}=0$ and $\mathrm{IPR}=1$ holding exactly; see Fig.~\ref{fig:spectra_evo_bhc}(a).

To further characterize the eigenstate-trajectory overlaps in the bosonic models, we studied the steady state purity, CM, and IPR.  The results are shown in Figs.~\ref{fig:indicators_bhd} and \ref{fig:indicators_bhc} for the Bose-Hubbard dimer and chain, respectively. Similarly to the XXZ chain discussed in the Main Text, we find a strong correlation between these three quantities. We also verified the lower bound on the IPR provided by the steady-state purity.

\section{Non-negativeness of the center of mass}

In this appendix, we prove that for a (pure) quantum trajectory, the CM defined in Eq.~(\ref{eq:CM_def}) (computed from $p_{\alpha,m}$) is non-negative.
To this end, we consider the following two-stage protocol. An initial pure state (at time $t=0$) is evolved using the quantum-jump protocol up to time $t$, yielding a quantum trajectory $|\psi_m(t)\rangle$. Afterward, the state is evolved for an additional interval $\tau$ (i.e., up to time $t+\tau$) deterministically with the Liouvillian $\mathcal{L}$, generally becoming mixed and approaching the steady state. Let $\rho_t(0)=|\psi_m(t)\rangle\langle \psi_m(t)|$ denote the pure state at time $t$ and $\rho_t(\tau)$ its subsequent evolution under the Liouvillian, $d\rho_t(\tau)/d\tau=\mathcal{L}[\rho_t(\tau)]$. The purity $P_t(\tau)=\tr{\rho_t(\tau)^2}$ evolves as
\begin{align}
    \frac{dP_t(\tau)}{d\tau} %&= \frac{d}{d\tau} \tr{\rho_t(\tau)^2} 
    =2\,\mathrm{Tr}\left\{\rho_t(\tau) \frac{d\rho_t(\tau)}{d\tau}\right\}
    =2\,\tr{\rho_t(\tau) \mathcal{L} [\rho_t(\tau)]}.
\end{align}
To proceed, we expand $\rho_t(\tau)$ and $\mathcal{L}[\rho_t(\tau)]$ in the left and right eigenbases of $\mathcal{L}$, respectively. Denoting the overlaps of the right and left eigenstates with $\rho_t(\tau)$ by $c_{\alpha,t}(\tau)$ and $d_{\alpha,t}(\tau)$ ($p_{\alpha,t}(\tau)$ the respective quasiprobability) and setting $\tau=0$ yields:
\begin{align}
    \frac{1}{2}\left.\frac{d P_t(\tau)}{d\tau} \right|_{\tau=0} &= \sum_{\alpha\beta} \tr{d_{\alpha,t}(0) \ell_\alpha^\dagger \lambda_\beta c_{\beta,t}(0) r_\beta}
    \nonumber\\
    &= \sum_\alpha \lambda_\alpha p_{\alpha,t}(0) 
    = \mean{\lambda}\,\text{CM}_t(\tau=0), \label{eq:purity_cm}
\end{align}
where $\mathrm{CM}_t(\tau)$ is the CM defined from $p_{\alpha,t}(\tau)$. Since $P_t(\tau)\leq 1$ and $P_t(0)=1$, we find that $\left.dP_t(\tau)/d\tau\right|_{\tau=0}\leq 0$, which implies that $\text{CM}_t(0)\,\mean{\lambda} \leq 0$. 
With $\mean{\lambda} \leq 0$, this implies $\text{CM}_t(0) \geq 0$. Since $\text{CM}_t(\tau=0)=\text{CM}(t)$ as defined in the Main text, this concludes the proof.

\end{document}